\documentclass[useAMS, usenatbib, usegraphicx, twocolumn]{mn2e}

\input epsf
\usepackage{graphicx}
\usepackage{color}
\usepackage{amsfonts}
\usepackage[draft]{hyperref}

\newcommand{\adsurl}[1]{\href{#1}{ADS}}

\newcommand{\beq}{\begin{equation}}
\newcommand{\eeq}{\end{equation}}
\newcommand{\barr}{\begin{eqnarray}}
\newcommand{\earr}{\end{eqnarray}}

\newcommand{\lesssim}{\mathrel{\hbox{\rlap{\lower.55ex\hbox{$\sim$}} \kern-.3em \raise.4ex \hbox{$<$}}}}
\newcommand{\gtrsim}{\mathrel{\hbox{\rlap{\lower.55ex\hbox{$\sim$}} \kern-.3em \raise.4ex \hbox{$>$}}}}

\title[Search for PAHs in the Perseus molecular cloud]{Search for PAHs
in the Perseus molecular cloud with the Green Bank Telescope}

\author[Ali-Ha\"{\i}moud, P\'erez, Maddalena \& Roshi]{
       Yacine Ali-Ha\"{\i}moud$^{1, 2}$\thanks{yacine@jhu.edu},
       Laura M.~P\'erez$^{3}$\thanks{L.~M.~P\'erez is a Jansky Fellow of
         the National Radio Astronomy Observatory}, 
       Ronald J.~Maddalena$^{4}$ and
       D.~Anish Roshi$^{5}$\\
${1}$ Department of Physics and Astronomy, Johns Hopkins University,
Baltimore MD 21218\\
${2}$ School of Natural Sciences, Institute for Advanced Study,
Einstein Drive, Princeton, NJ 08540, USA \\
${3}$ National Radio Astronomy Observatory, P.O. Box O, Socorro, NM
87801, USA \\
${4}$ National Radio Astronomy Observatory,  P.O. Box 2. Green Bank,
WV 24944, USA\\
${5}$ National Radio Astronomy Observatory, Green Bank and Charlottesville, VA 22903, USA}

\begin{document}

\date{\today}

\pagerange{\pageref{firstpage}--\pageref{lastpage}} \pubyear{2014}
\pagenumbering{arabic}
\label{firstpage}

\maketitle

\begin{abstract}

Polycyclic Aromatic Hydrocarbons (PAHs) are believed to be the
small-size tail of the interstellar carbonaceous dust grain
population. Their vibrational emission is the most widely accepted source of the aromatic near-infrared
features, and their rotational radiation is a likely explanation for
the dust-correlated anomalous microwave emission (AME). Yet, no
individual interstellar PAH molecule has been identified to date. It
was recently recognised that quasi-symmetric planar PAHs ought to have a
well identifiable comb-like rotational spectrum, and suggested
to search for them in spectroscopic data with matched-filtering
techniques. We report the results of the first such search, carried
out with the Green Bank Telescope, and targeting the star-forming
region IC348 in the Perseus
molecular cloud, a known source of AME. Our observations amounted to
16.75 hours and spanned a 3 GHz-wide band extending from
23.3 to 26.3 GHz. Using frequency switching, we achieved a sensitivity of 0.4 mJy per 0.4 MHz channel ($1
\sigma$). The non-detection of comb-like spectra allowed us to
set upper bounds on the abundance of specific quasi-symmetric PAH
molecules (specified uniquely by their moments of inertia) of
approximately 0.1\% of the total PAH abundance. This bound generically applies
to PAHs with approximately 15 to 100 carbon atoms. \\

\end{abstract}


\section{Introduction}

It is nowadays widely accepted that Polycyclic Aromatic Hydrocarbons
(PAHs) are ubiquitous in the interstellar medium (ISM). These large and
very stable molecules constitute the small-size tail of the
carbonaceous dust grain population. They are important actors in the
thermal and chemical balance of the ISM \citep{Tielens_2008}. The
mid-infrared (IR) aromatic features at 3-20 $\mu$m, main
evidence for PAHs, are believed to result from their vibrational
emission following stochastic heating by ultraviolet (UV) photons. Their strength implies that a few percent of
the interstellar carbon is locked in these molecules
\citep{Allamandola_1989}. The dust-correlated ``anomalous microwave
emission'' (AME, \citealt{Leitch_1997}), at a few tens of GHz, is most
likely rotational emission from PAHs (``spinning dust'',
\citealt{DL98_long}). Finally, electronic transitions in PAHs could be
the explanation for some of the diffuse interstellar bands (DIBs)
observed in the optical and UV \citep{Salama_2011}. 

Despite its three decades of existence, the PAH hypothesis still lacks an
ultimate piece of evidence: the unambiguous identification of specific
PAH molecules in space. Such a detection would bring a definitive end to
lingering disputes over the nature of the carriers of mid-IR aromatic
features \citep{Kwok_2011, Li_2012, Kwok_2013, Yang_2013}. The search
for specific PAHs in space is, however, a challenging task. The aromatic IR bands, cornerstone of the PAH
hypothesis, arise from nearest-neighbour C--H
and C--C vibrations; as a result, they cannot be used to identify specific PAH
molecules in space. In fact, the observed mid-IR spectra can be
reproduced by mixtures of a few tens of different PAH
species, but the exact composition of the mixtures is not critical
\citep{Rosenberg_2014}.

The far-infrared regime probes the lowest energy modes
corresponding to the overall bending of the skeleton, and could provide a
successful avenue to identify specific PAHs \citep{Mulas_2006}. However these modes carry only a small fraction of
the total radiated energy and their contrast with the underlying
continuum is expected to be low \citep{Tielens_2008},
which has impeded their detection so far. UV and optical spectroscopy has also
been used to search for specific PAHs through their electronic
transitions, which has allowed to set upper bounds on the abundance of a
few specific PAHs, but not lead to any unambiguous detection yet
(\citealt{Gredel_2011, Galazutdinov_2011}; see however
\citealt{Iglesias_2008, Iglesias_2010, Iglesias_2012}
for tentative
detections of naphtalene and anthracene towards the Perseus molecular cloud).

Rotational spectroscopy is a prime tool in the identification of
interstellar molecules and the measurement of their abundances, and has
obviously been suggested for interstellar PAHs, though rather
timidly \citep{Hudgins_2005, Tielens_2008, Hammonds_2011}. To date however, only the
bowl-shaped corannulene molecule (C$_{20}$H$_{10}$) has been
searched for with this technique, without success \citep{Lovas_2005,
  Thaddeus_2006, Pilleri_2009}. The reasons for the scarcity of works
using or even suggesting this technique are twofold. First, there could be a large number
of PAH molecules and their individual spectra could be lost in the
forest of their combined emission. Second, PAHs are large, generally triaxial
molecules, unlike the very special polar and symmetric corannulene
molecule. One may therefore a priori expect their rotational emission to be complex and diluted
over a very large number of weak lines.

Recently, one of us argued that the prospects of rotational
spectroscopy of PAHs may not be so bleak (\citealt{YAH_2014}, hereafter
AH14). First, one can reasonably
expect that a few highly symmetric and stable ``grand PAHs'' could be
more fit to survive the harsh ISM conditions and as a result are over-abundant \citep{Tielens_2013}. Secondly, even though triaxiality is
unavoidable if a planar PAH is to have a permanent dipole moment, the
degree of asymmetry need not be large, and the resulting rotational
spectrum may still remain relatively simple. In particular,
quasi-symmetric nitrogen-substituted PAHs\footnote{To avoid clutter we
  shall still call PAH any molecule that derives from an actual PAH (in the rigorous
  sense of the term) by minor modifications, such as the substitution of C
  or H atoms by other atoms, or the attachment of a radical on the
  periphery. }, strongly polar and
likely common in the ISM \citep{Hudgins_2005}, ought to have a
well-identifiable ``comb''-like rotational spectrum. In addition to
being less diluted, the simple shape of the spectrum allows for the
use of matched-filtering techniques, resulting in an enhanced
sensitivity. Furthermore, one can undertake blind searches of a
priori unknown molecules by scanning over comb spacings, without the need for precise and extensive
quantum-mechanical calculations or laboratory measurements of their
rotational constants.

\begin{figure}
\includegraphics[height = 50 mm]{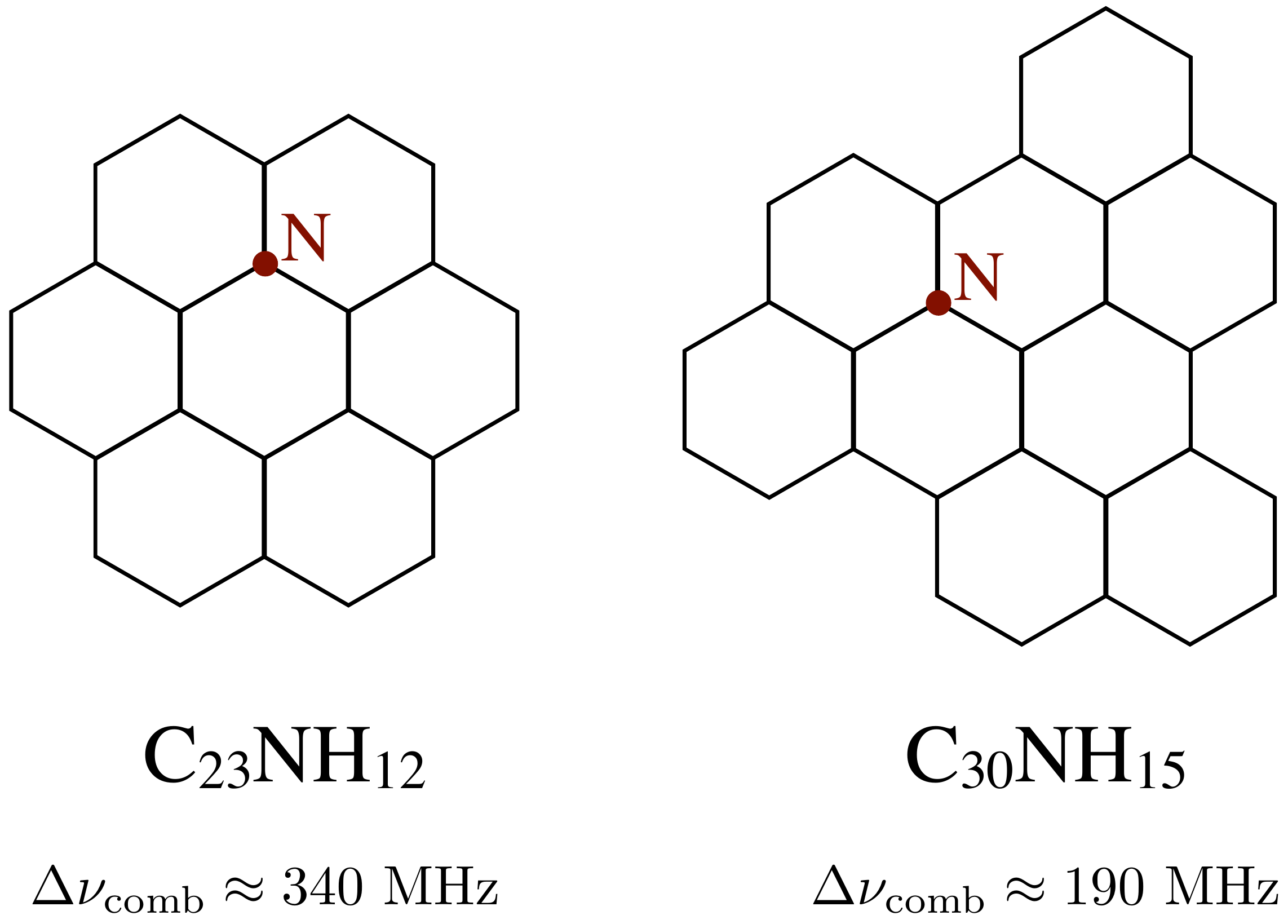}
\caption{Examples of nitrogen-substituted symmetric PAHs which remain
  quasi-symmetric after substitution, in the sense that their two
  smallest moments of inertia differ by less than a few percent. Such
  molecules are the target of this search. $\Delta \nu_{\rm comb}$ indicates the
  approximate line spacing of their comb-like rotational spectrum (as estimated in AH14).} \label{fig:example} 
\end{figure}

\begin{figure}
\includegraphics[height = 47 mm]{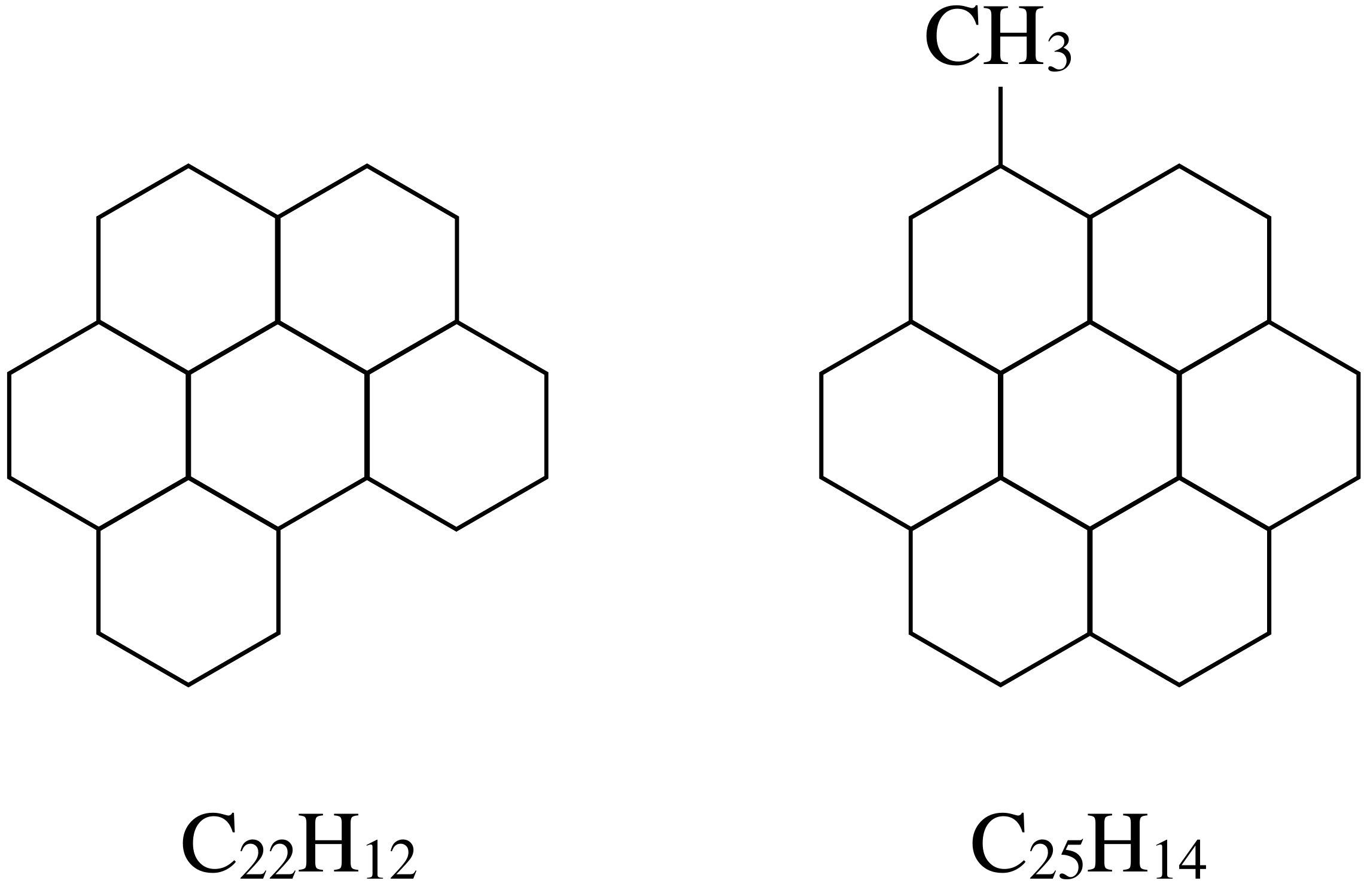}
\caption{Examples of PAHs or derivatives that \emph{do not qualify} as
  quasi-symmetric, as the fractional difference between any two of their
  moments of inertia is larger than a few percent. The rotational
  spectrum of such large triaxial molecules is complex and diluted
  among many lines, and we do not
  attempt to search for them.} \label{fig:notsym} 
\end{figure}

In this paper we describe our search for quasi-symmetric interstellar
PAHs (we show in Figs.~\ref{fig:example} and \ref{fig:notsym} explicit
examples of PAHs that do or do not
qualify as quasi-symmetric). We rely on rotational spectroscopy and matched
filtering, using data taken with the Robert C.~Byrd Green Bank Telescope
(GBT)\footnote{The project ID is GBT/14A-504}. We have targeted the
Perseus molecular cloud, a known region of AME \citep{Tibbs_2013},
which, if it is indeed spinning dust radiation, is just the collective
rotational emission from PAHs. Our observations have reached a
sensitivity of 0.4 mJy per 0.4 MHz channel across a 3 GHz total
bandwidth. This sensitivity was not enough to detect any comb-like
spectrum in our data, but we could set upper bounds on the fractional abundance
of specific quasi-symmetric PAHs, assuming the observed AME in Perseus
is primarily due to spinning dust emission. 

This paper is organised as follows. In Section \ref{sec:theory} we
summarise the calculation of AH14 for the rotational emission of
quasi-symmetric planar PAHs and describe the theoretical model and
assumptions we rely on to set upper bounds. Our data reduction method
is described in detail in Section \ref{sec:reduction}. Section
\ref{sec:analysis} describes our search for combs in our reduced and
calibrated data. We derive upper bounds on abundances of individual
PAHs in Section \ref{sec:bounds}, and conclude in Section \ref{sec:conclusion}.

\section{Theoretical background} \label{sec:theory}

In this section we briefly summarise the findings of AH14.


Any rigid rotor is characterised by three moments of inertia $I_1 \leq
I_2 \leq I_3$, to which correspond three rotational constants $A_i
\equiv \hbar/(4 \pi I_i)$, with dimensions of frequency. The
spectrum of a quasi-classical (angular momentum $L \gg \hbar$)
triaxial rotor is in general very complex, with a large number of
allowed transitions, resulting in a dilution of the radiated energy
per transition. 

AH14 considered the restricted class of \emph{planar},
\emph{quasi-symmetric} rotors, having in mind nitrogen-substituted symmetric
PAHs. For a classical planar rotor, with the third axis perpendicular
to the plane, $I_1 + I_2 =
I_3$. However, this relation is never exactly satisfied for a
quantum-mechanical rotor, and there is always a small but non-zero \emph{inertial
  defect} 
\beq
\delta \equiv 1 - \frac{I_1 + I_2}{I_3}.
\eeq
A perfectly symmetric rotor would have $A_1 = A_2$, and the small departure from
this relation is quantified by the asymmetry parameter
\beq
\epsilon \equiv \frac{A_1 - A_2}{2 A_3}.
\eeq
The rotational state of such quasi-symmetric rotors is characterised by two quantum numbers: the total
angular momentum $J$ and its projection $K$ along the axis of
(quasi)symmetry of the grain, with $-J \leq K \leq J$. Assuming the electric
dipole moment is in the plane of the grain, the
strongest transitions are the $(J, K) \rightarrow (J-1, K \pm
1)$. To lowest order in $\epsilon$ and $\delta$, the frequency
$\nu_{JK}$ of the $\Delta K = -1$ transition (or that of the $(J, -K)
\rightarrow (J, -K+1)$ transition) is given by (AH14)
\barr
\nu_{JK} &=& A_3(4 J - 2K +1) \nonumber\\
&+& 4 A_3 \delta (J-K) + \frac{\epsilon^2}{4} A_3 (J+K) \left(\frac{J-K}{K}\right)^3. \label{eq:nuJK}
\earr
The power radiated in this transition is
\beq
P_{JK} = \frac{\mu^2}{6 c^3} (2 \pi \nu_{JK})^4(1 + K/J)^2,
\eeq
where $\mu$ is the electric dipole moment. We see from Eq.~(\ref{eq:nuJK}) that for small asymmetries and
inertial defects, the families of transitions with constant
$2J - K \equiv J_0$ are clustered around $\nu_{J_0} = A_3 (2 J_0 +
1)$, reached exactly for $J = K = J_0$. Therefore the spectrum has the appearance of a \emph{comb}, with each
``tooth'' (labelled by $J_0$) being a cluster of quasi-coincident
lines. If the probability distribution
of $J$ peaks around $J_0$, and given that the power radiated is
proportional to $(1 + K/J)^2$, most of the power is radiated for
$\frac23 J_0 \leq J \leq J_0$, corresponding to $J/2 \leq K \leq J$,
and giving a characteristic relative spread for each
cluster\footnote{AH14 gave a larger analytic estimate of the width of
  each cluster (Equation 29) because the branch $-J \leq K \leq -J/2$ was also
  included there. However, at constant probability distribution for $J$ and $K$, there is
  significantly less power radiated in this branch (by two orders of magnitude) than in the $J/2
  \leq K \leq J$ branch, and the estimate we provide here is more accurate.}
\beq
\frac{\Delta \nu}{\nu_{J_0}} \approx \frac23 \delta +
\frac{\epsilon^2}8.
\eeq
For $\nu_{J_0} \approx 25$ GHz, we find that the width of each cluster
is less than our resolution of 0.4 MHz as long as $\delta \lesssim 3
\times 10^{-5}$ and $\epsilon \lesssim 0.01$. AH14 estimated the asymmetry resulting from a single substitution of a
carbon atom by a nitrogen atom in coronene and circumcoronene. The
resulting values of $\epsilon$ are all below 0.01, except for the
outermost substitution site in coronene, for which $\epsilon \approx
0.018$. We therefore expect that for most quasi-symmetric
nitrogen-substituted PAHs, the
asymmetry is small enough to preserve the ``comb''-like aspect of the
spectrum when observed with 0.4 MHz resolution. It is difficult to
estimate the inertial defect of large molecules from first principles. Based on a few measured or computed values for small PAHs, AH14
argued that one can expect it to be of the order of a few parts in
$10^5$. Provided this estimate is correct, the inertial defect should
also be small enough to preserve a ``comb'' appearance. However, we caution
that this is not a robust estimate, and it would be valuable to have accurate computations of
the inertial defect of large PAHs. 

Finally, we point out that corannulene also has a comb-like spectrum,
resulting from transitions $\Delta J = -1, \Delta K = 0$ and an out-of
plane electric dipole moment. The frequencies of these transitions are
$\nu_J = J \times 1.019685$ GHz \citep{Lovas_2005}. 
 
\section{Observations}

\subsection{Choice of target}

We have chosen to target the Perseus molecular cloud complex, which has been observed thoroughly at radio
and infrared frequencies, and is a well-known source of AME. Using data from the \emph{Planck} satellite complemented by ancillary
data from several other experiments, the Planck Collaboration (2011) fit the spectrum of
the Perseus cloud with a free-free, dust and AME
components. They find an AME contribution of 33.4 $\pm$ 4.5 and 31.5
$\pm$ 4.4 Jy at 22.8 and 28.5 GHz, respectively, with a FWHM
resolution of 1.12 degree. Tibbs et al. (2010) used the Very Small
Array (VSA) to carry out observations of the region with higher spatial
resolution at 33 GHz, finding a peak flux of 0.2 Jy per $7'$ beam. This
is consistent with the Planck Collaboration's results, accounting for
the loss of power due to the large-scale continuum being resolved out
by the VSA interferometer. Recently, \cite{Tibbs_freefree_2013} used
GBT observations at 1.4 and 5 GHz to constrain the amount of free-free
emission in the Perseus cloud. They find that free-free emission
cannot contribute more than 27\% (3$\sigma$) of the 33 GHz emission observed by
the VSA, confirming its ``anomalous'' nature. Finally,
\cite{Tibbs_2013} observed the Perseus cloud at 16 GHz with the Arcminute
Microwave Imager (AMI), and found a significant spatial correlation of the
emission with infrared emission from small dust grains, strengthening
the hypothesis that the AME is indeed due to small spinning dust grains.

We pointed the GBT towards the strongest peak of AME in the Perseus
molecular cloud, as measured by \cite{Tibbs_2013}. The coordinates of the AME peak are RA:
03:44:28.4, Dec: +32:04:42.1 (C.Tibbs, private communication).
 
\subsection{Summary of the observations}

Our observations were carried out in five sessions during the month of
March 2014, each divided in several scans of
approximately half an hour. At the start of each scan, and towards a
calibrator near our science target, we measured the telescope focus
and pointing offsets in azimuth and elevation. We did not pursue a
out-of-focus holography to measure the surface large-scale errors,
given the low frequency of our observations. We employed the
high-frequency part of the KFPA receiver together with the GBT
spectrometer, to obtain the widest possible bandwidth at high spectral
resolution. Our observations spanned 3 GHz of bandwidth, from 23.3 to
26.3 GHz, by setting up four partially overlapping 800 MHz-wide widows. Each window was configured to provide 2048 channels with a spectral resolution of 0.390625 MHz. This spectroscopy setup allowed for a single beam on the sky and two circular polarisations to be observed.

The entire set of observations amounted to a total telescope time of
16.75 hours, including 8.7 hours of actual exposure on source and 2.2
hours of exposure on two blank fields off-source. The off-source
observations were used for null tests in our analysis. Our observing setup and
data calibration (described in the next section) allowed us to reach a sensitivity of 0.4 mJy per 0.4 MHz channel.

\section{Data reduction}\label{sec:reduction}

\subsection{Properties of the raw data}

\begin{figure*}
\includegraphics[height = 50 mm]{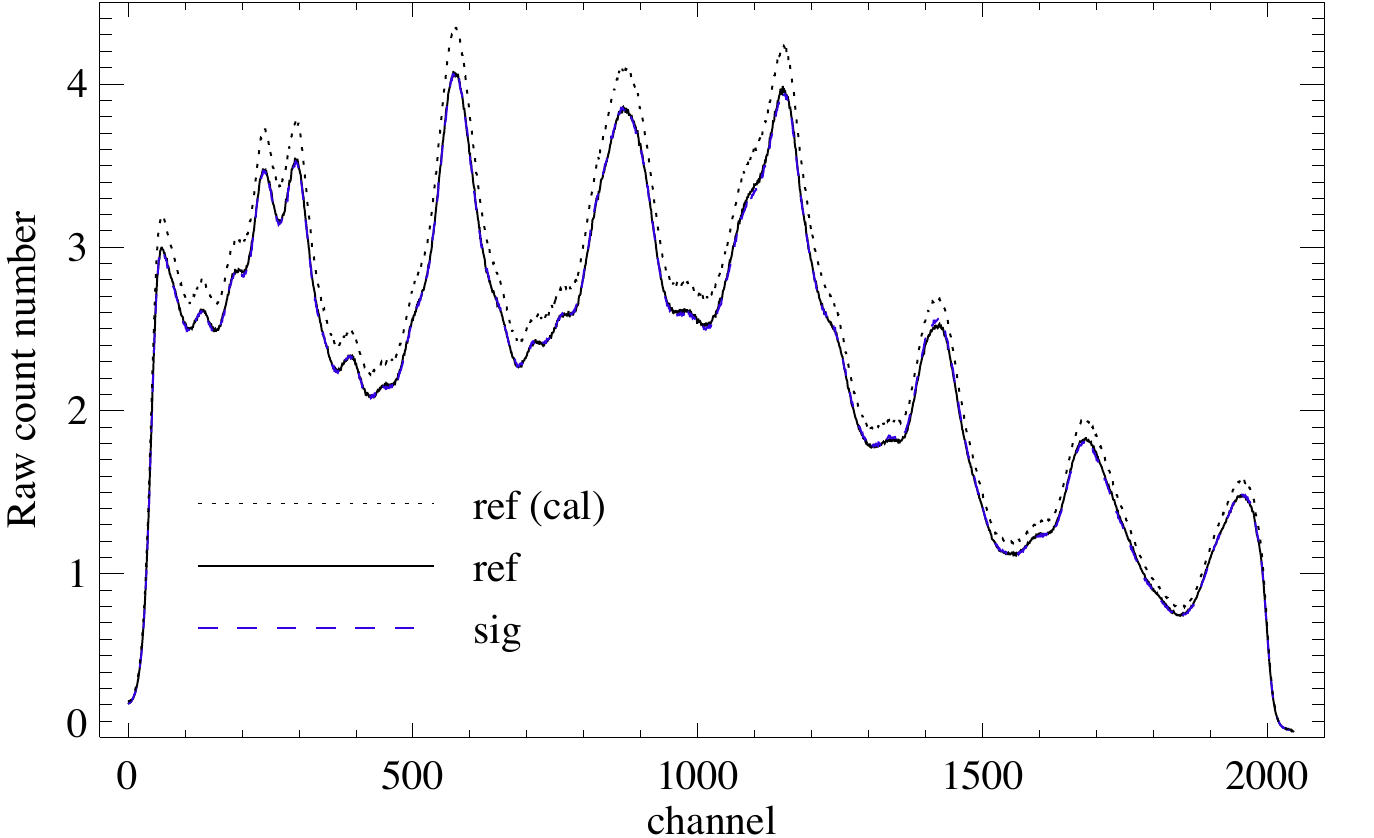}
\includegraphics[height = 50 mm]{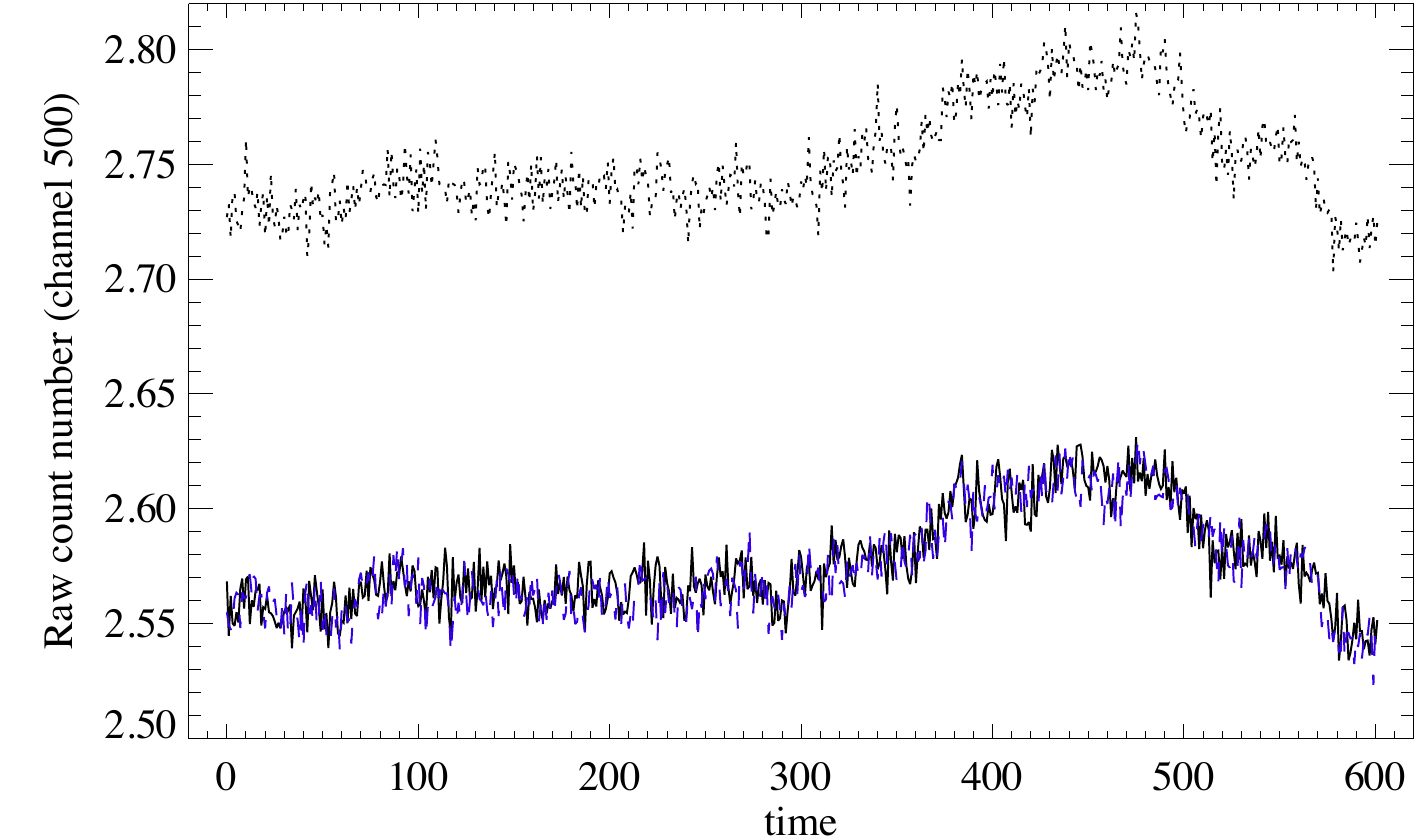}
 \caption{Raw count numbers for the lowest 800 MHz band of the first
   scan, in one polarisation state. \emph{Left panel:} single time
   sample as a function of channel number. \emph{Right panel:}
   Evolution of channel 500 as a function of time (a unit increment on the
   $x$-axis corresponds to approximately 2 seconds of actual time).
} \label{fig:counts} 
\end{figure*}

The raw data is given in terms of \emph{count number} $P(n, t)$, as a
function of channel number $n$ and time sample $t$. There are $N_{\rm
  ch} = 2048$ channels per 800 MHz band. Each time sample corresponds
to an integration of $2$ seconds, split between the four phases
described below. The sky frequency
(in the local standard of rest) corresponding to channel $n$ is
denoted by $\nu_n$. There are two
frequency-shifted phases: a ``reference'' phase and a
``signal'' phase, differing in sky frequency by a constant shift $\Delta \nu_{\rm fs} =
10$ MHz:
\beq
\nu_n^{\rm sig} = \nu_n^{\rm ref} + \Delta \nu_{\rm fs}.
\eeq
Each of these two phases is itself divided into two sub-phases,
with a calibrating noise diode switched off or on
(subscript ``cal'' for the latter). We denote the raw count numbers for these four
phases by $P^{\rm ref}, P^{\rm sig}, P^{\rm ref}_{\rm cal}$ and
$P^{\rm sig}_{\rm cal}$, respectively. We show a sample of the raw
data in Figure \ref{fig:counts}.

The raw count numbers are related to the antenna temperature
$T_a(\nu)$ of the line emission we are aiming to measure as follows: for each phase ph
= ref, sig, we have
\barr
P^{\rm ph}(n, t) &=& G^{\rm ph}(n, t) \left[T_a(\nu_n^{\rm ph}) + T_{\rm
    sys}(\nu_n^{\rm ph}, t)  \right],\\
P^{\rm ph}_{\rm cal}(n, t) &=& G^{\rm ph}(n, t)
\left[T_a(\nu_n^{\rm ph}) + (T_{\rm
    sys}+ T_{\rm cal})(\nu_n^{\rm ph}, t) \right], ~~~~~\\
G^{\rm ph}(n, t) &=& G_{\rm IF}(n, t) G_{\rm RF}(\nu_n^{\rm ph}, t).
\earr
In these equations, the intermediate-frequency gain $G_{\rm IF}$ is
the system gain after the local oscillator (where the frequency
switching is done) in the optical path. It is
identical for the reference and signal phases, up to small variations
that may occur during the 1-second lag between them. The radio-frequency gain $G_{\rm RF}$ represents
the system gain before the local oscillator, and may differ slightly
between the two phases. $T_{\rm sys}$ denotes the
total system temperature, including the continuum
emission from the observed region. $T_{\rm cal}$ is the temperature
of the calibrating noise diode, assumed to be well
known and constant over each one of the four 800 MHz sub-bands (and
slightly different for each polarisation state), see
\cite{Bryerton_2011} for a description. The assumed calibrator temperature pairs for each band are, in Kelvins: $\hat{T}_{\rm cal} =
(3.243,  3.285), (3.733, 3.841), (4.341, 4.429), (4.791,
4.871)$. 

The gain has mostly large-scale fluctuations in the frequency domain due to
standing waves in the optical path. The largest-scale fluctuations are
well visible in Fig.~\ref{fig:counts}. On
the other hand, ideal thermal noise is uncorrelated from one
channel to another and fluctuate with equal power on all scales, with
a variance given by the radiometer equation
\beq
\langle \delta T_{\rm sys}^2 \rangle = \frac{T_{\rm sys}^2}{\Delta
  \nu_{\rm res}
\Delta t}, 
\eeq
where $\Delta t \approx 0.4$ s is the exposure time per phase. 
 
These behaviours are better illustrated in Fourier space. We define our
Fourier convention as follows: for a function $P(n)$ of channel
number, the Fourier transform $\tilde{P}(u)$ is defined for $u = 0,
... N_{\rm ch}/2$ as
\beq
\tilde{P}(u) \equiv \frac{1}{N_{\rm ch}} \sum_{n = 0}^{N_{\rm ch}-1}
P(n) \exp\left(- i \frac{2 \pi}{N_{\rm ch}} u n\right).
\eeq
The power spectrum of the raw count number is obtained by averaging $2
|\tilde{P}(u; t)|^2$ over the few hundred integrations of a given
scan. It gives the contribution to the variance of the count number
from fluctuations with period $\Delta n = N_{\rm ch}/u$ channels.

\begin{figure}
\includegraphics[height = 50 mm]{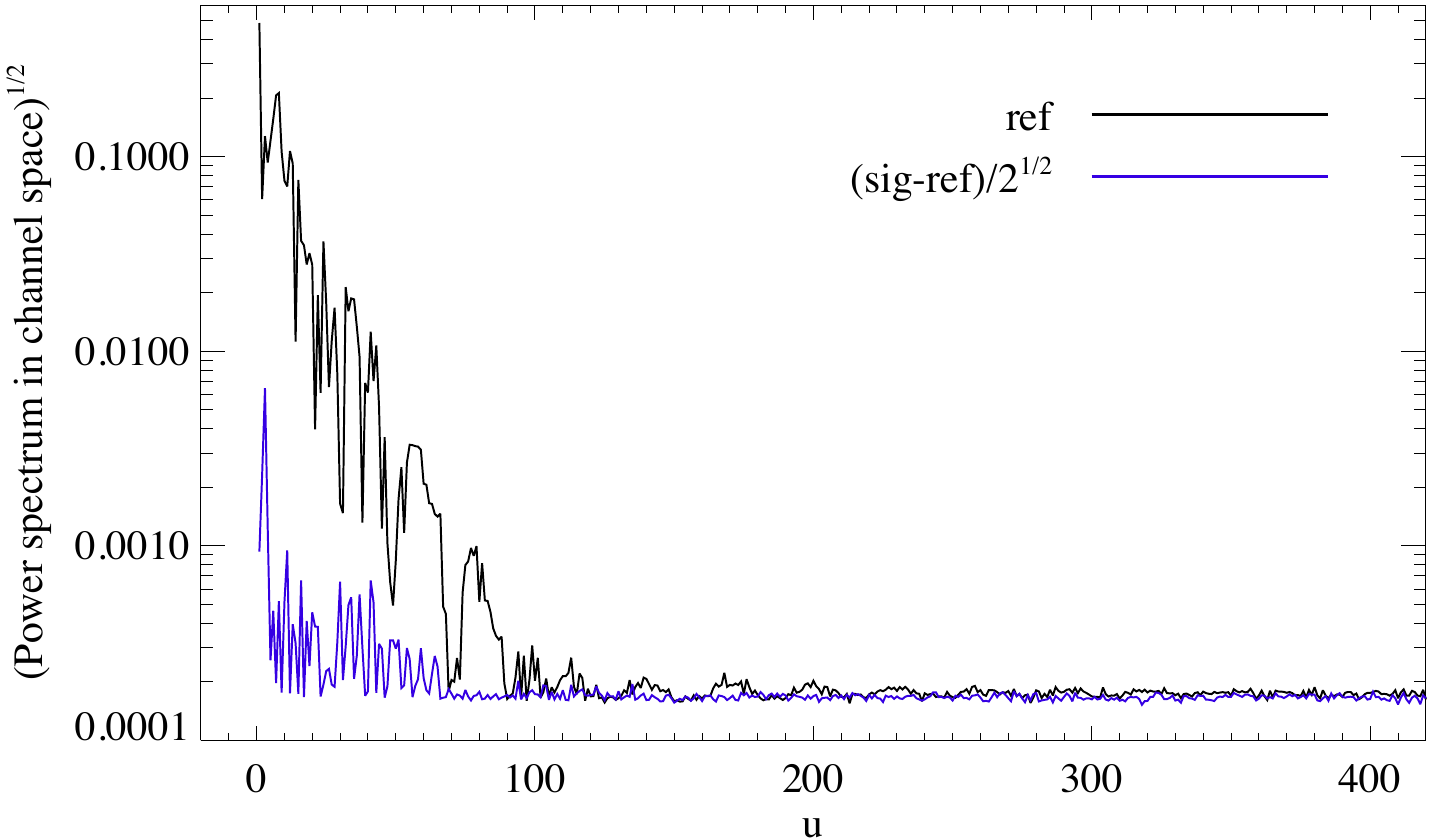}
 \caption{Square root of the power spectrum of the count number for a
   given scan for the reference and signal phases separately, and for
   their normalised difference. The variable $u$ corresponds to
   fluctuations with period $\Delta n = N_{\rm ch}/u$ channels.} \label{fig:pow_spec} 
\end{figure}

The upper (black) line in Fig.~\ref{fig:pow_spec} shows the power spectrum of
the reference phase. We see that fluctuations on small frequency scales have a flat power spectrum, as expected from
thermal noise. On large scales $u \lesssim 100$, non-thermal gain fluctuations
dominate the power spectrum. 
The lower (blue) line shows the power spectrum of
the difference $(\textrm{sig} - \textrm{ref})/\sqrt{2}$. We see that
this combination eliminates a substantial part of large-scale gain
fluctuations. However, there are still significant non-thermal fluctuations due to the
imperfect cancellation of the RF gain between the two phases.


Armed with this basic understanding of the properties of the data, we now
describe our data reduction method.

\subsection{Data reduction method}

\subsubsection{Elimination of glitches}

Our first step in the data analysis is to identify glitches in the time
evolution of each channel. We first compute the local average of the count number
by smoothing it over 100 channels in time (we use a Savitsky-Golay
smoothing kernel). We then discard any data that differs from this
smoothed count number by more than 7 times the expected rms thermal noise
fluctuation. We inspect visually all scans that show a large number of
flagged data points. We find three distinct cases of large gain
fluctuations: \\
$(i)$ Transient radio-frequency interference, due to
passing satellites. These events are brief (typically lasting a
few tens of seconds) and are visible in both polarisations. Even
though these glitches are localised in frequency they are so large in amplitude
that they affect a broad range of channels. We simply discard all
the channels of the affected band during the duration of the glitch.\\
$(ii)$ Large gain instabilities, often limited to a single
polarisation, and lasting for a whole scan. We discard the whole scan from the affected polarisation
and bands.\\
$(iii)$ Detector resonances, taking the form of large glitches in
evenly spaced channels ($n = 255 + i \times 256$), and in both the reference and signal
phases. We just discard the corrupted channels when this is the
case. 

\subsubsection{Estimation of the system temperature} \label{sec:gain}

\begin{figure}
\includegraphics[height = 50 mm]{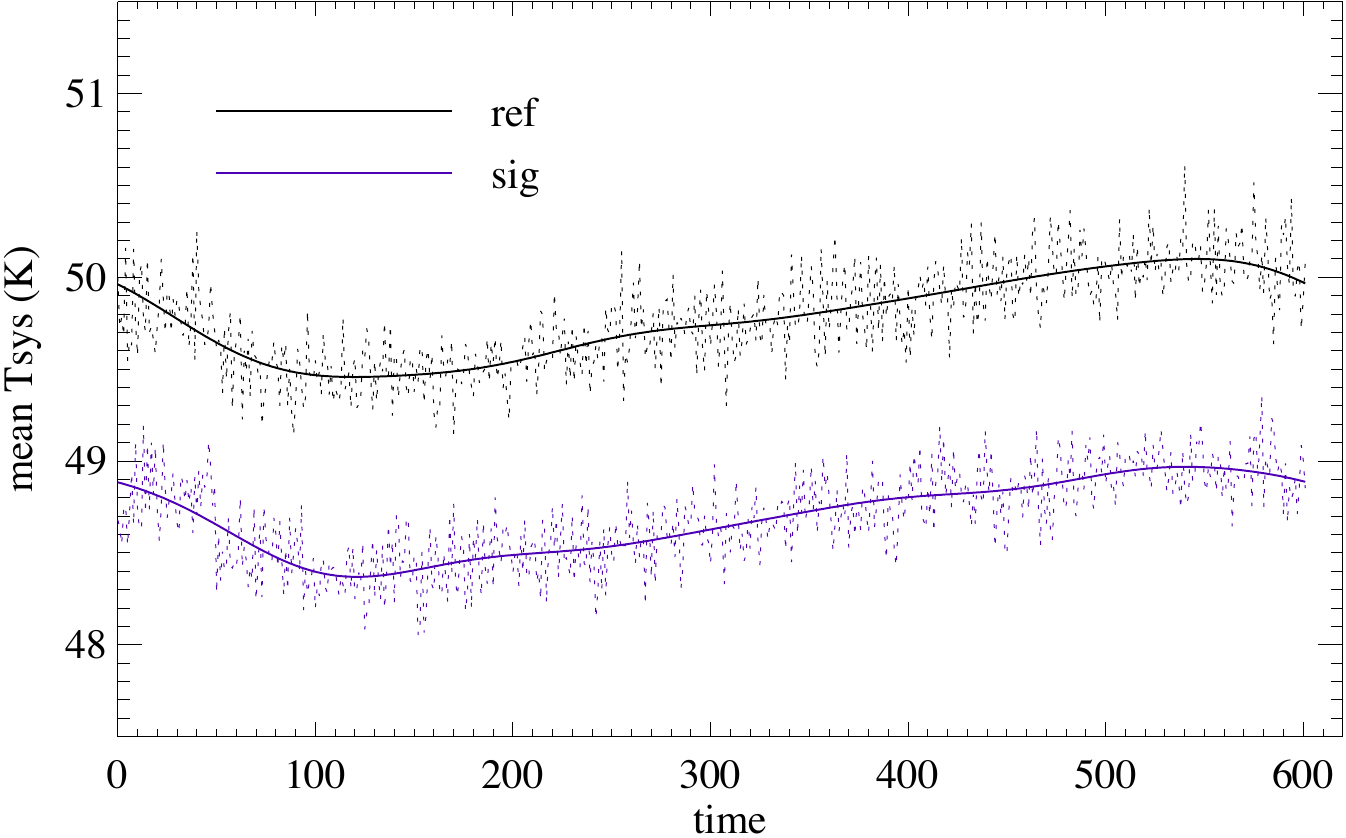}
 \caption{Estimate of the system temperature for the reference (upper,
   black lines) and signal (lower, blue lines) phases, averaged
   over the innermost 90\% of the band. The solid lines show the slow,
   non-thermal evolution of the system temperature with time. The
   $\sim 2\%$ difference between the two phases is systematic across
   all scans and bands and is likely due to a slight
   mis-calibration. It has no effect on our analysis.} \label{fig:mean_Tsys} 
\end{figure}

We estimate the system temperature for each of the two phases as follows:
\beq
\hat{T}_{\rm sys}^{\rm ph}(n, t) \equiv \hat{T}_{\rm cal} \frac{P^{\rm
  ph}(n, t)}{P^{\rm ph}_{\rm cal}(n, t) - P^{\rm ph}(n, t)}.
\eeq
This estimate is noisy, with a fractional error of the order of
$\delta T_{\rm sys}/T_{\rm cal}$. Since we only have estimates of the
mean calibrator temperature across each band, we can only hope to
estimate a mean system temperature as well. We therefore average
$\hat{T}_{\rm sys}$ over the innermost 90\% of the band. The resulting
mean system temperature (typically of the order of 50 K) is shown with dashed lines in Fig.~\ref{fig:mean_Tsys} as a
function of time. Besides remaining thermal fluctuations of order
$\delta T_{\rm sys}/T_{\rm cal}/\sqrt{N_{\rm ch}}$, there are also
larger systematic variations on timescales of a few minutes. We keep
the first five modes of the time Fourier transform of $\hat{T}_{\rm
  sys}(t)$ as our final estimate for the mean system temperature. This
is shown with solid lines in Fig.~\ref{fig:mean_Tsys}.

\subsubsection{Estimation of the antenna temperature}\label{sec:Ta}

Our basic approach to eliminate the gain relies on baseline fitting in
Fourier space. We denote by $\overline{P}(n, t)$ the count number smoothed in
channel space and $\delta P \equiv P - \overline{P}$ the
complementary, rapidly fluctuating part. The smoothing is done by keeping the Fourier modes $u \leq u_0
\equiv 200$ only, using a hyperbolic tangent with width $\Delta u =
20$ as a truncating function. Our estimate of the antenna temperature for each phase is then
\barr
\hat{T}_a^{\rm ph} &\equiv& \hat{T}_{\rm sys}^{\rm ph} \frac{\delta P^{\rm ph}}{\overline{P}^{\rm
    ph}},\\
\hat{T}_{a, \rm cal}^{\rm ph} &\equiv& (\hat{T}_{\rm sys}^{\rm ph} +\hat{T}_{\rm cal}) \frac{\delta P^{\rm ph}_{\rm cal}}{\overline{P}^{\rm
    ph}_{\rm cal}}, 
\earr
with noise of variance $T_{\rm sys}^2/\Delta t \Delta \nu_{\rm res}$
and $(T_{\rm sys} + T_{\rm cal})^2/\Delta t \Delta \nu_{\rm res}$,
respectively. We average these two estimates with an inverse-variance
weighting, and then similarly average the result over all integrations of
each scan. 

\begin{figure*}
\includegraphics[height = 50 mm]{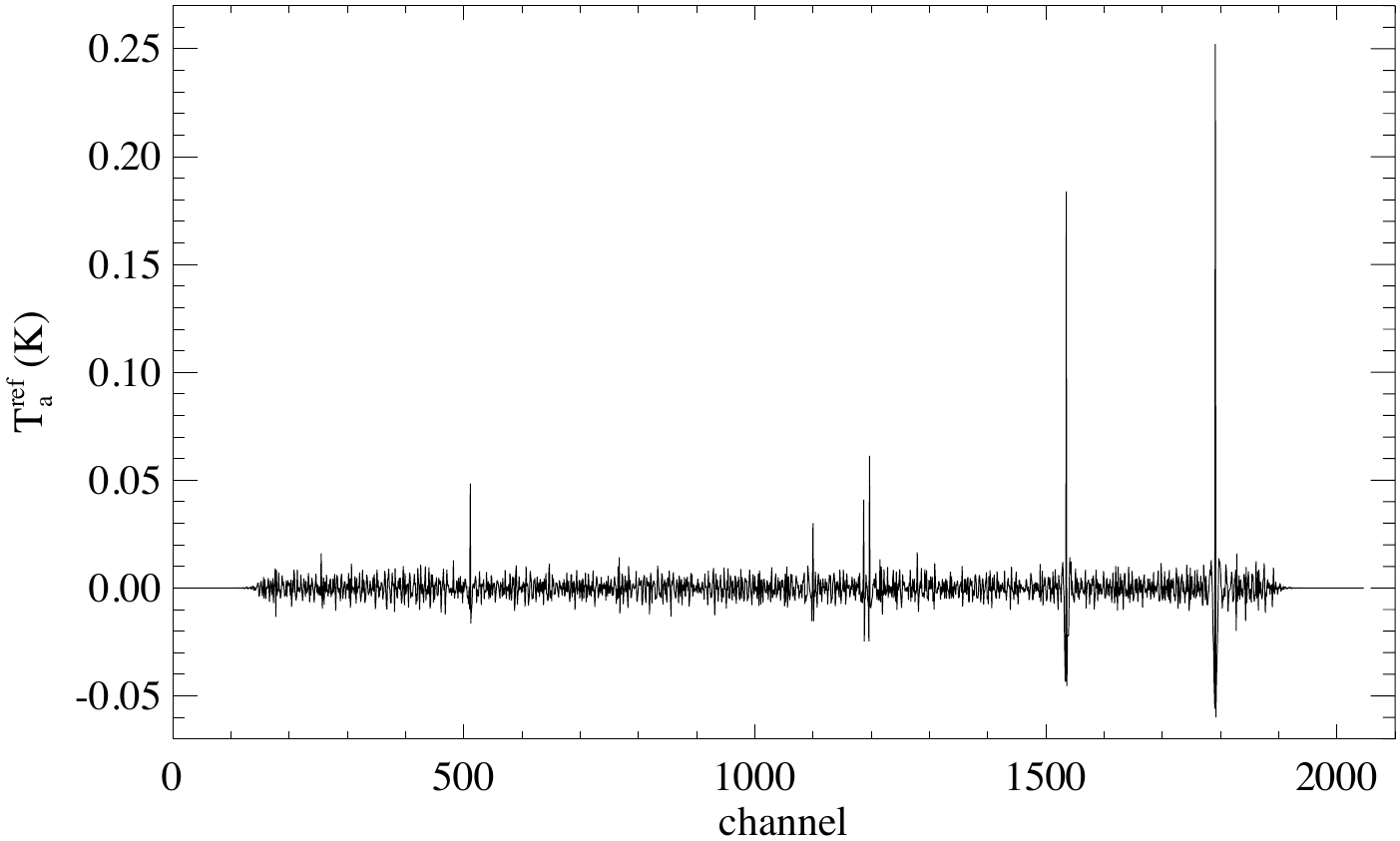}
\includegraphics[height = 50 mm]{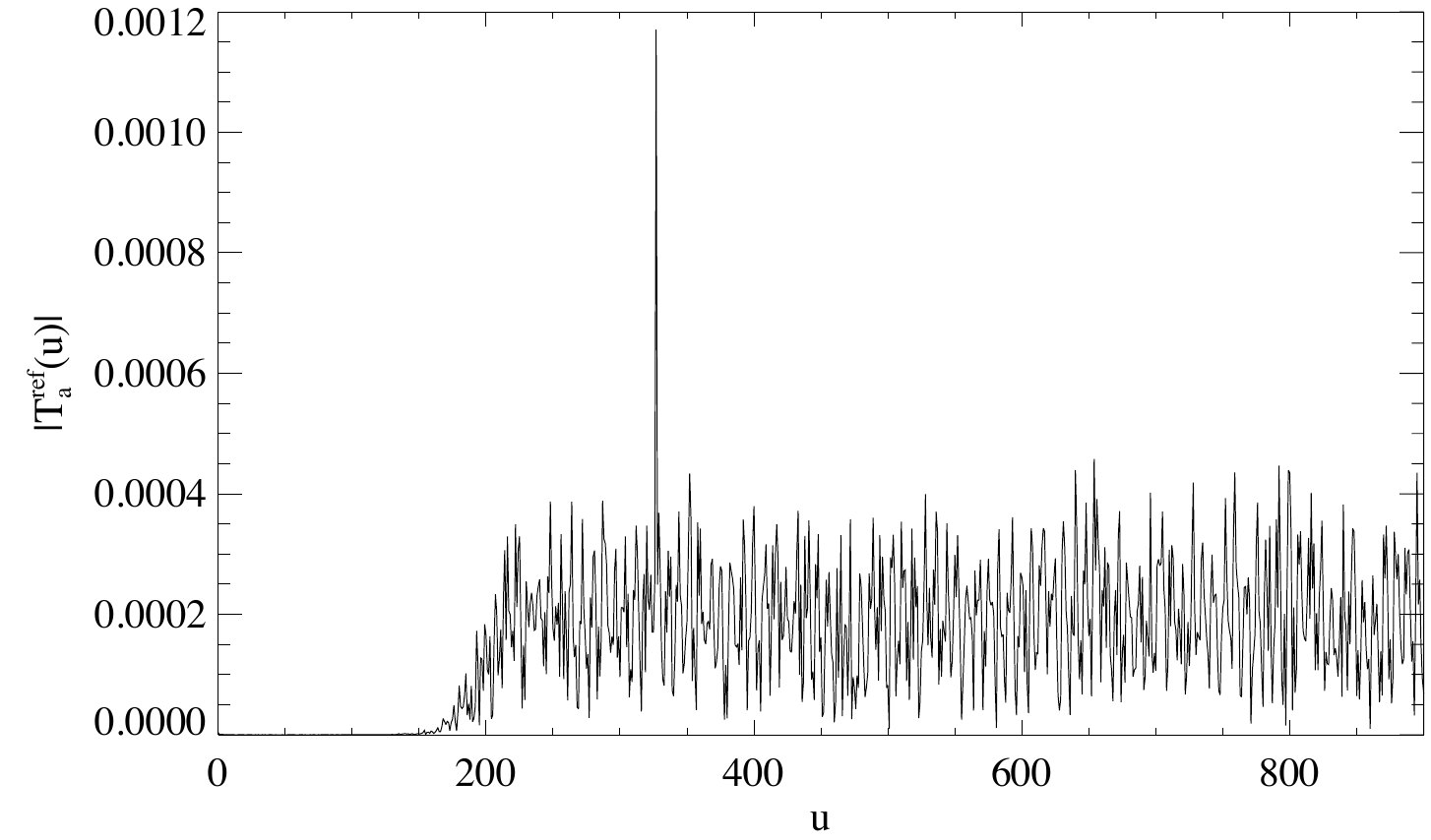}
 \caption{Spectrum of the reference phase averaged over one scan, in
   channel space (left panel), and in Fourier space (right panel), see
 description in Sec.~\ref{sec:Ta}.} \label{fig:Ta_ref} 
\end{figure*}

We show the resulting spectrum for the reference phase in
Fig.~\ref{fig:Ta_ref}, in channel space and in Fourier space, after
truncating it outside of the 90\% innermost channels with a smooth truncating function.
The left plot clearly shows large spikes at a few evenly spaced
   channels $(n = 255 + i \times 256$), which
   translate into a dense comb in Fourier space. These channels
   correspond to the peaks of the standing waves in the gain. In this
   case theses channels did not show any anomalous behaviour in the
   time evolution of the gain and were therefore not flagged in the
   preliminary analysis. The Fourier transform
 also shows a large spike at $u = 327$, corresponding to standing
 waves with very short period $\Delta \nu = 2.44$ MHz. These spikes
 are clearly undesirable high-frequency gain fluctuations that were
 not eliminated by the smoothing.

\begin{figure*}
\includegraphics[height = 50 mm]{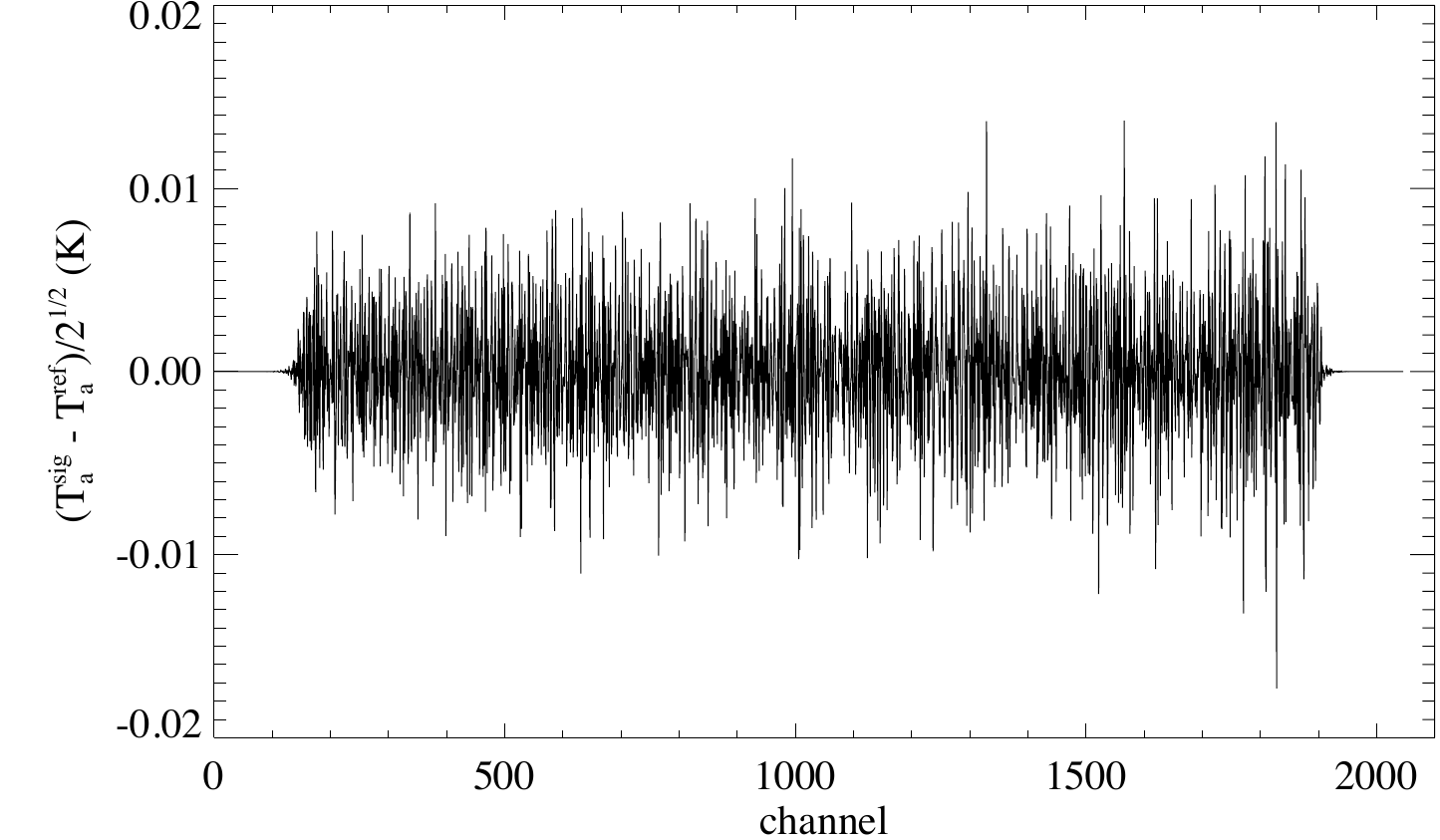}
\includegraphics[height = 50 mm]{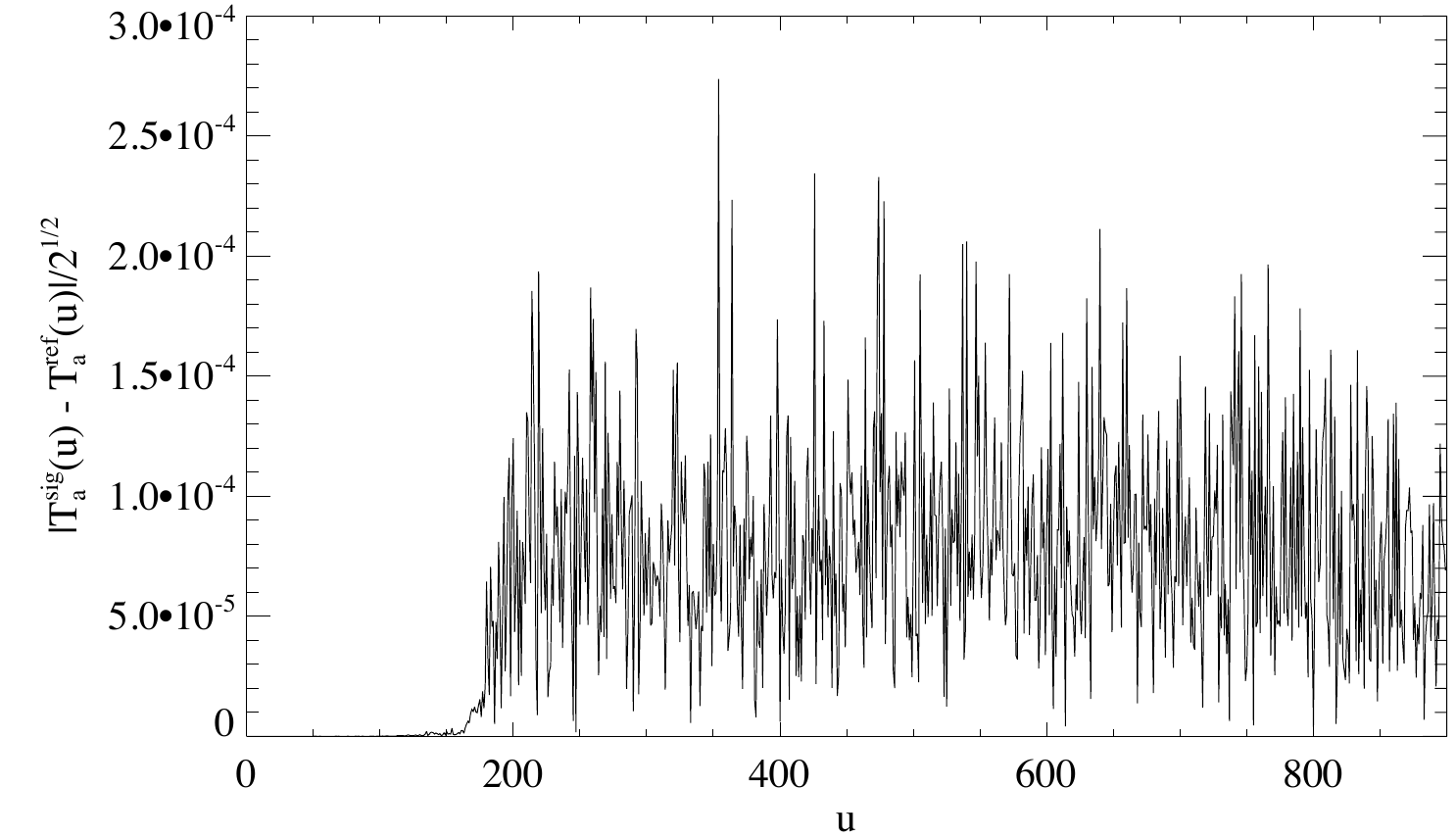}
 \caption{Differential spectrum $(T_a^{\rm sig} - T_a^{\rm ref})/\sqrt{2}$ in
   channel space (left panel), and in Fourier space (right
   panel). This combination eliminates the undesirable high-frequency
   gain fluctuations seen in each phase alone (compare to Fig.~\ref{fig:Ta_ref}).} \label{fig:Ta_sig_ref} 
\end{figure*}

Since these features are due to the IF gain, they are cancelled by
frequency-switching, i.e. subtracting the reference from the signal phase. This is illustrated
in Fig.~\ref{fig:Ta_sig_ref}, where we can see that the difference
spectrum $T_a^{\rm sig} - T_a^{\rm ref}$ is consistent with thermal
noise (besides the truncation at $u \lesssim 200$).

The difference spectrum is then averaged over the two polarisation
states and all scans of a given session, with a minimum variance weighting.

\begin{figure}
\includegraphics[height = 50 mm]{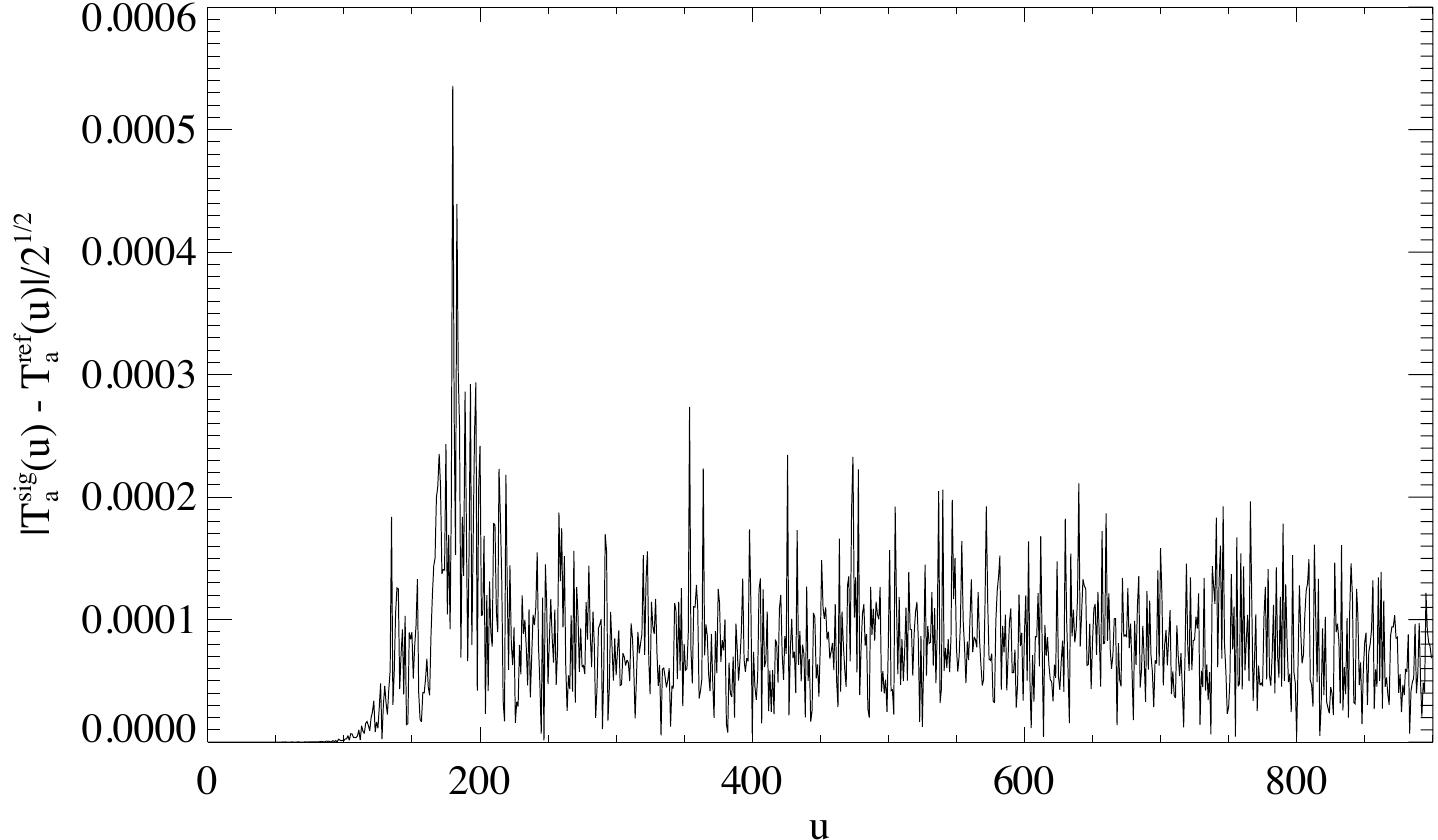}
\caption{Incomplete elimination of gain fluctuations when using a too
  gentle cutoff (in this case $u_0 = 150$ instead of
  200). Modes $u\approx 150-200$ are significantly contaminated by RF
  gain fluctuations, which are not cancelled by frequency switching.} \label{fig:u0_150} 
\end{figure}

While it is critical to eliminate gain contamination from our
estimate of the antenna temperature, it is also in our interest to keep as
many modes as possible as usable data. Indeed, partial truncation of
Fourier modes at $u_0$ leads to a reduction of the amplitude of the signal, by an amount
$2 u_0/N_{\rm ch} \approx 20\%$. The variance of the noise is also
reduced (and it becomes correlated), but only by the square root of the latter fraction. We have
checked that it is not possible to use a truncation value smaller than
$u_0 = 200$ without suffering from RF gain contamination at $u \sim
100-200$, which cannot be eliminated by frequency switching. This is
illustrated in the spectrum shown in Fig.~\ref{fig:u0_150}, estimated
using a cutoff at $u_0 = 150$
instead of 200. Note that the RF gain fluctuations were not visible in
Fig.~\ref{fig:pow_spec}, which showed the average power spectrum of a
single integration. Averaging over a few hundred integrations
decreases the white noise plateau, to the point where it becomes
dominated by the high-frequency tail of the RF gain fluctuations at $u
\lesssim 200$. 

\subsubsection{Calibration}

In order to convert the estimated antenna temperature to a flux, we
have observed the quasar 3C48. \cite{Perley_2013} provide the following fit for its spectral
density flux, as of January 2012:
\barr
&&\log_{10}(S_{\nu}[\textrm{Jy}]) = 1.3324 - 0.769 \log_{10}(\nu_{\rm
  GHz}) \nonumber\\
&&-0.195[\log_{10}(\nu_{\rm GHz})]^2 + 0.059 [\log_{10}(\nu_{\rm GHz})]^3.
\earr
This translates to mean fluxes of 1.148, 1.114, 1.082 and 1.052 Jy,
respectively, when averaging over the central 80\% of each one of our
four bands. 

We have observed 3C48 once per session (except for the first session),
and obtained the mean brightness temperature in each band by differentiating the
mean system temperature measured on source and off source. Dividing by the known
fluxes, we find sensitivities between 1.3 and 1.85 K/Jy, varying
between observing sessions and bands\footnote{This analysis did not
  account for the change in atmospheric opacity from one session to
  another nor for the change in pathlength through the atmosphere
  during each session. We have computed the total atmospheric
  opacity and found that the resulting error in calibration is at most
  10\%. Such corrections can therefore be safely ignored.}

For the first session, where the calibrator was not observed, we estimate the conversion factor $T/S_{\nu}$ by its average
over the four remaining sessions.

The final estimate of the flux (or rather of $\Delta S(n)$, the
difference between signal and reference phases) is obtained by an inverse-variance
averaging of our five observing sessions.

\subsubsection{Additional cleaning}

After obtaining the reduced and calibrated spectra, we clean the data
further as follows:

$(i)$ We truncate the band edges where the gain is low and the
estimated flux is noisy: typically, we remove 50-100
channels (20-40 MHz) on either side of each band.

$(ii)$ We eliminate regions showing strong non-thermal
contamination. The highest frequency band is the most affected, with
very large contamination near 25.69 GHz, affecting nearly 40\% of the
band. This contamination is most likely due to a gain issue: it
is present in both polarisations but with significantly different amplitudes.

$(iii)$ We re-smooth the data in Fourier space and discard modes $u
\leq 220$, corresponding to fluctuations with period larger than 9.3
channels or 3.6 MHz. The highest-frequency band also shows non-thermal
contamination for very high frequency modes ($u \approx N_{\rm
  chan}/2$). These high-frequency fluctuations
are due to the large feature near 25.69 GHz, and we filter them out.

\subsubsection{Final reduced spectrum}

\begin{figure*}
\includegraphics[height = 50 mm]{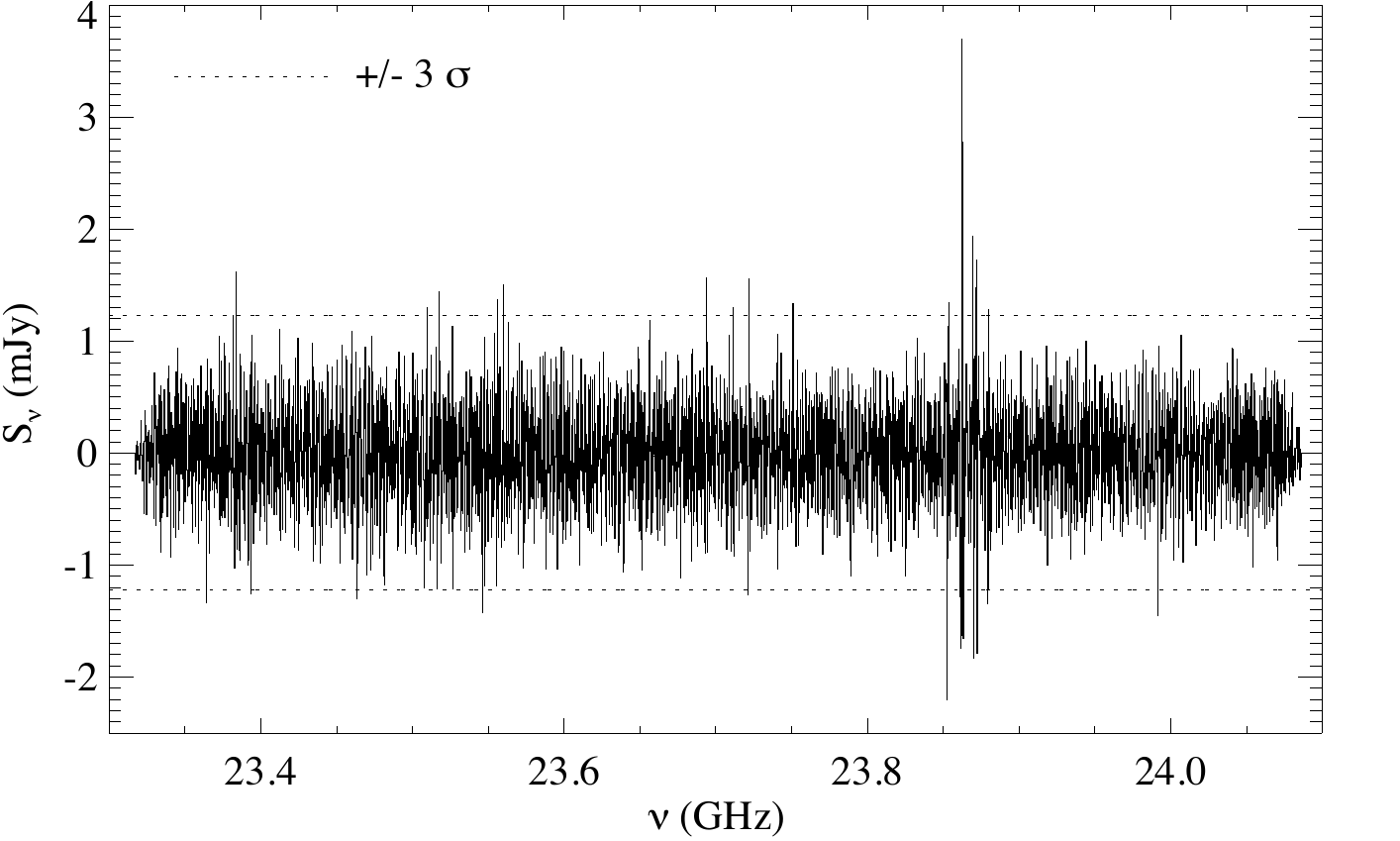}
\includegraphics[height = 50 mm]{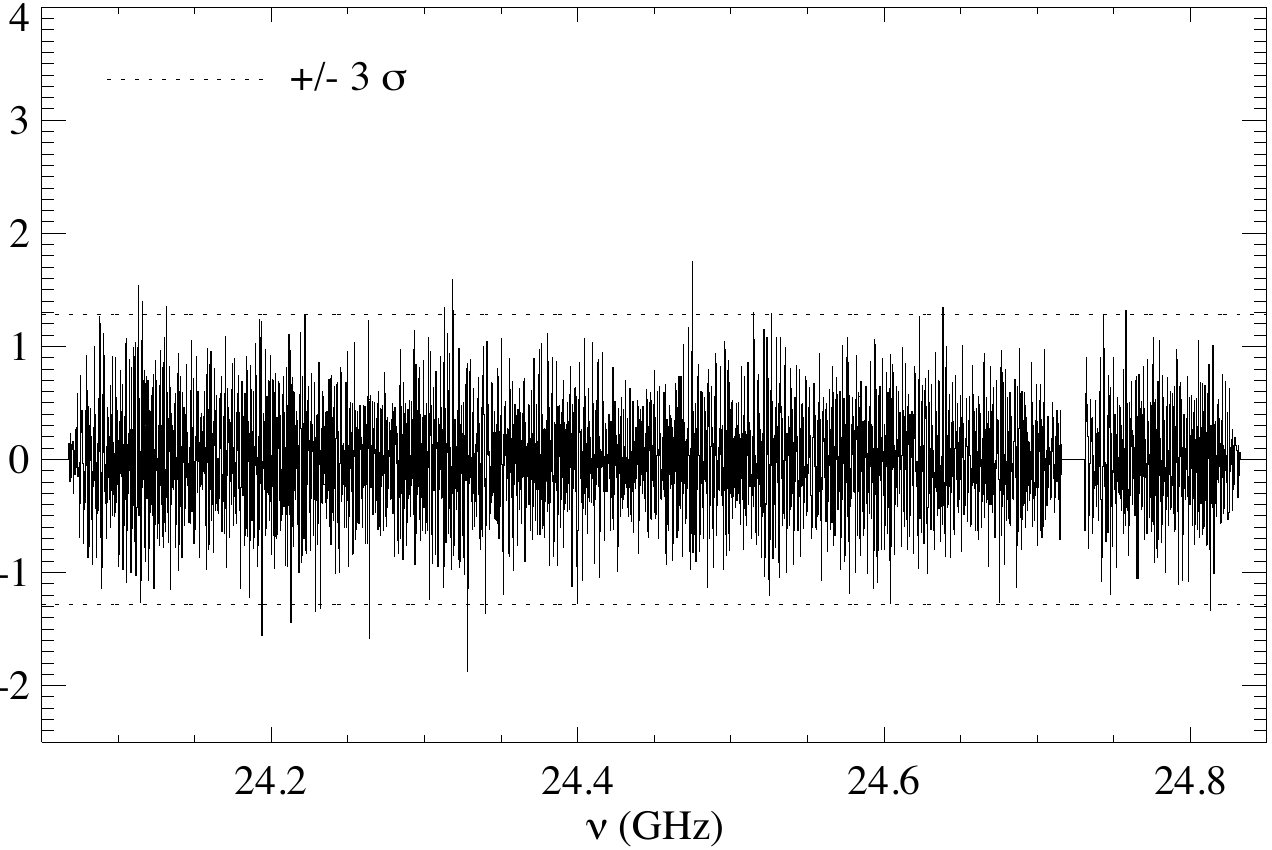}
\includegraphics[height = 50 mm]{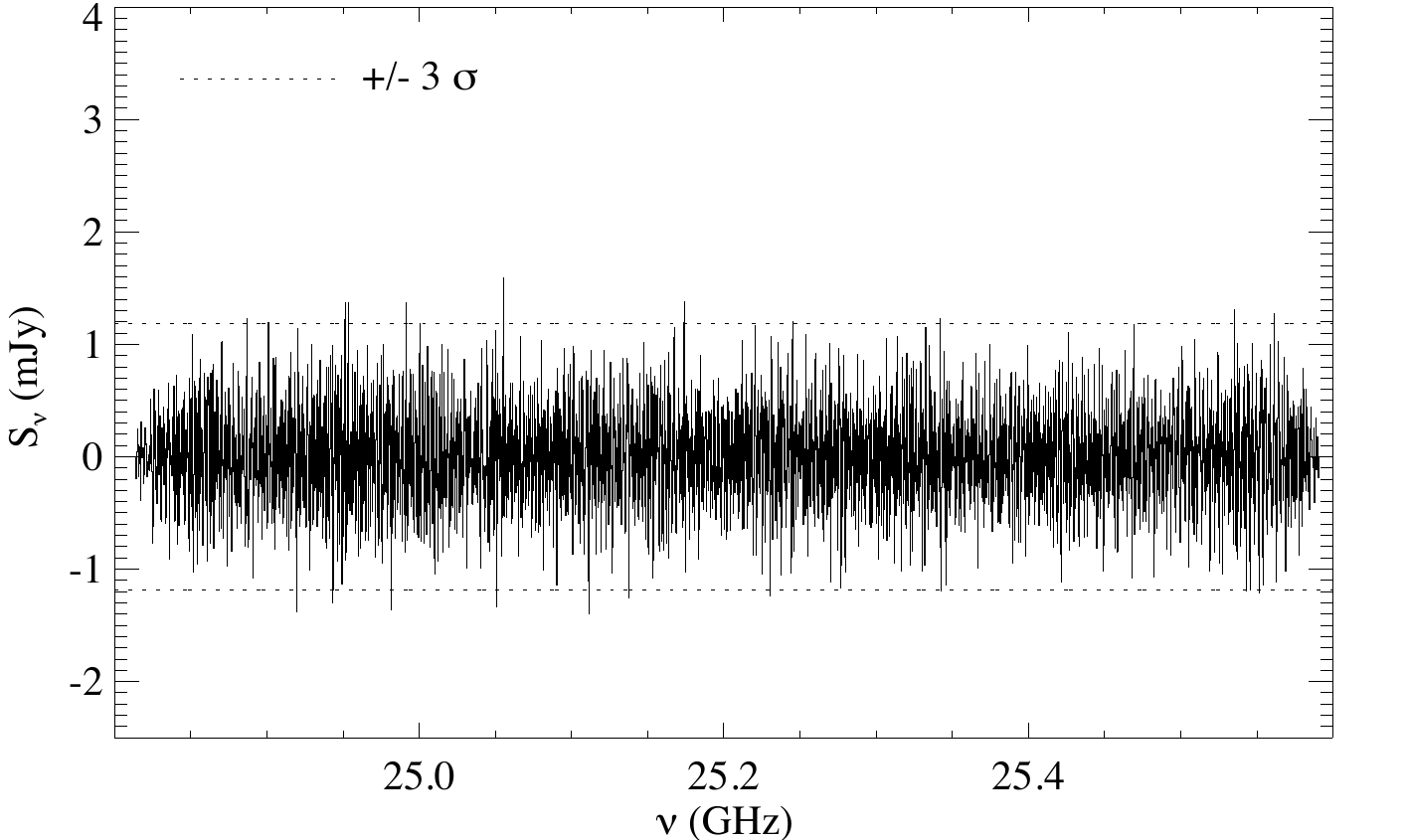}
\includegraphics[height = 50 mm]{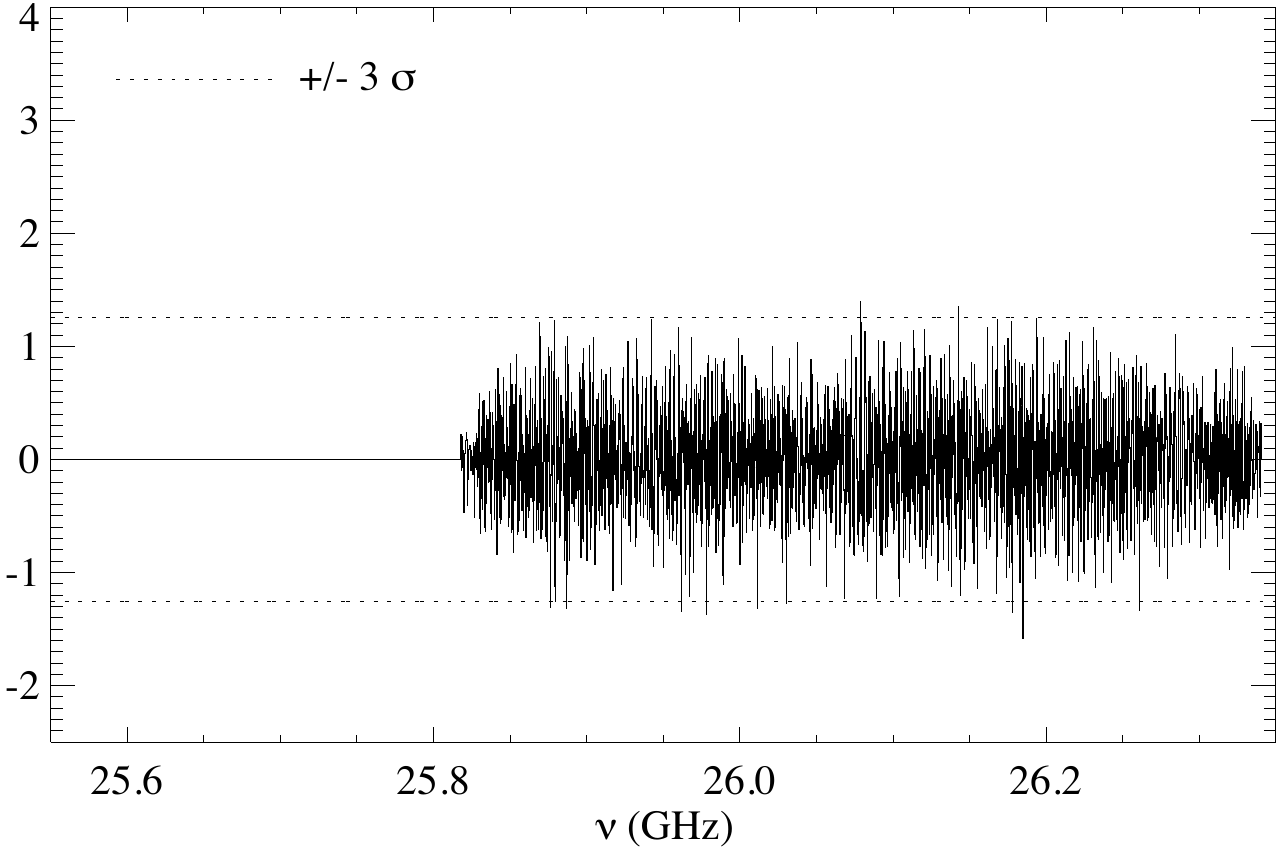}
\caption{Final reduced and calibrated flux for each one of our four 800
  MHz bands --
  specifically, this is an estimate of $S(\nu) + [S(\nu + 10~ \rm
  {MHz}) - S(\nu - 10~ \rm{MHz})]/2$. The dashed lines show the $\pm 3
\sigma$ bands, at approximately $\pm 1.2$ mJy. The blanked regions are discarded data due to large contamination.
The strong feature at $\nu = 23.86$ GHz is likely due
to ozone emission and is masked during the PAH line search. The
negative satellite features at $\pm 10$ MHz are due to the frequency switching.} \label{fig:flux} 
\end{figure*}

The final estimate of the flux $\hat{S}(\nu)$ is obtained by the combination
\beq
\hat{S}(\nu) = \frac{\Delta S(n^{\rm sig}) - \Delta S(n^{\rm ref})}2,
\eeq
where $n^{\rm sig}$ is the channel such that $\nu = \nu_n^{\rm sig}$
(within a channel width), and
similarly for $n^{\rm ref}$, so that 
\beq
\hat{S}(\nu) = S(\nu) - \frac{S(\nu + \Delta \nu_{\rm fs}) + S(\nu -
  \Delta \nu_{\rm fs})}{2} + \textrm{noise}.
\eeq
We show our final reduced spectrum $\hat{S}(\nu)$ in Fig.~\ref{fig:flux}.

We have reached noise levels of 0.39 - 0.43 mJy ($1\sigma$) per
channel, in each of the four 800 MHz bands. The
spectrum is consistent with thermal noise. 

\begin{figure}
\includegraphics[height = 50 mm]{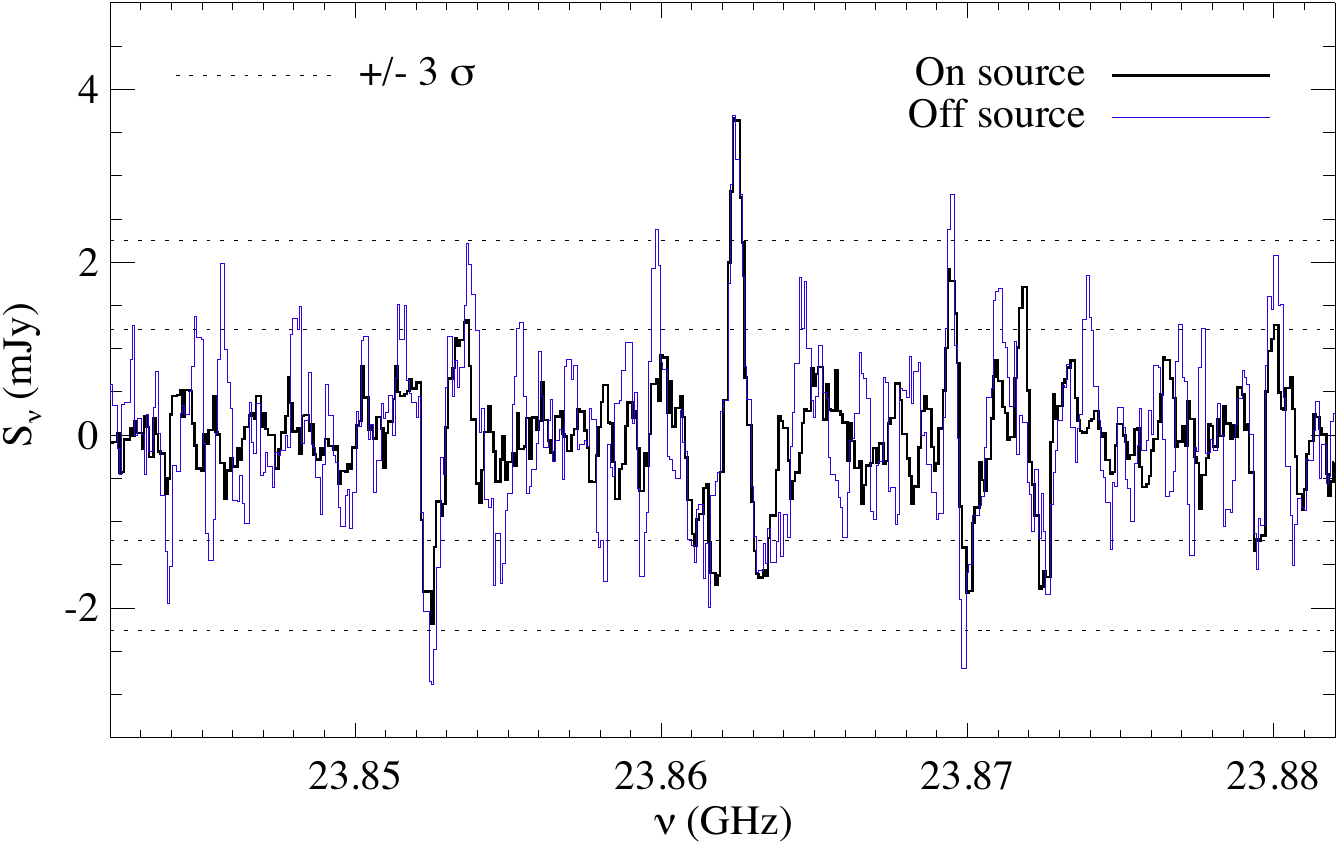}
\caption{Detail of the spectrum around 23.86 GHz, for our observations
on source (black, thick line) and off-source (blue, thin lines). The
dashed lines show the $\pm 3 \sigma$ boundaries for the two spectra
(the noise for the off-source observations is approximately twice
larger than that of the on-source observations). The ozone line
appears clearly in both spectra as a trong positive feature at $\nu \approx 23.862$ GHz and two
negative satellites at $\pm 10$ MHz due to frequency switching. The
immediate negatives surrounding the main feature (and positives
surrounding the negative satellites) result from high-pass filtering in
Fourier space when reducing the data. Finally, a strong feature also appears at 23.87 GHz, in
both spectra, and we also mask it for the PAH line search.} \label{fig:ozone}
\end{figure}

We detect an intense line at 23.862 GHz, with a flux of
3.6 mJy. This line also appears with high signal-to-noise in the spectrum of the blank field off
source (see Fig.~\ref{fig:ozone}), and therefore does not originate
from the source. We attribute it to
atmospheric ozone emission \citep{Gora_1959}. We also detect a weaker but still
intense feature at 23.87 GHz, in both on and off-source spectra (see
Fig.~\ref{fig:ozone}). We mask both of these features in the PAH line
search. We checked that besides these two features, no
strong line in the on-source spectrum also appears at more than 2 sigma in
the off-source spectrum.

\section{Data analysis}\label{sec:analysis} 

\subsection{Search for PAH combs}

We search for combs in the data of the form $\nu_J = (J + 1/2) \Delta
\nu_{\rm comb}$, where $J$ is an integer and $\Delta \nu_{\rm comb}$ is a fixed comb
spacing. The signal-to-noise ratio (SNR) for a comb with spacing $\Delta
\nu_{\rm comb}$ is simply obtained by summing the intensities at the expected
comb frequencies, and normalising by the noise rms fluctuation times
the square-root of the number of lines within the bandwidth. 

Two combs with spacings $\Delta \nu_1$ and $\Delta \nu_2$ are not differentiable if their
spacings are close enough, due to the finite resolution of the
data. The combs are distinguishable if $\Delta \nu_J = \nu_J \times
\Delta \log(\Delta \nu_{\rm comb}) \geq \Delta \nu_{\rm res}$, that is if 
\beq
\Delta \log(\Delta \nu_{\rm comb}) \gtrsim \frac{\Delta \nu_{\rm res}}{\nu}
\approx 1.5 \times 10^{-5}.
\eeq 
We use a step $\Delta \log(\Delta
\nu_{\rm comb}) = 10^{-6}$ to make sure we do not miss any
resonance.

\begin{figure}
\includegraphics[height = 55 mm]{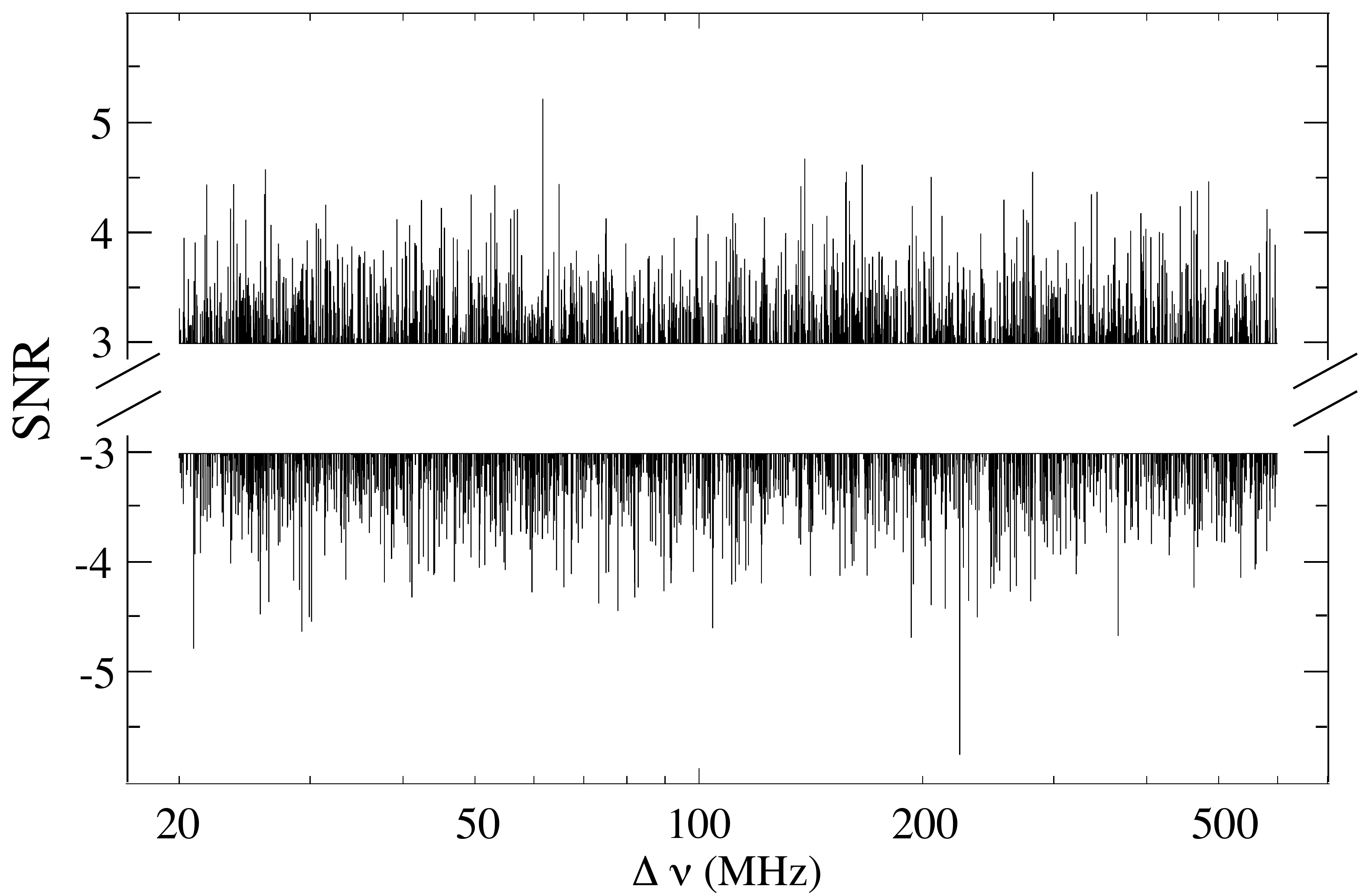}
\caption{Signal-to-noise ratio (SNR) as a function of comb spacing
  $\Delta \nu$. Given the very large number of comb spacings sampled
  (several millions), we only show the points with $|$SNR$| > 3$ for
  the sake of clarity. The number of comb
  spacings with $|$SNR$| > 4$ and $|$SNR$| > 5$ is consistent with what
  we find when carrying the same analysis on the off-source spectrum.} \label{fig:snr}
\end{figure}

We have computed the SNR for combs with spacings
$\Delta \nu_{\rm comb} = 20$ MHz to 600 MHz, corresponding to PAHs
with approximately 15 to 100 carbon atoms, and including the expected
spacing for coronene and circumcoronene ($\Delta \nu_{\rm comb} \approx
340$ and $70$ MHz, respectively). The result is shown in
Fig.~\ref{fig:snr}. We find that it is
consistent with a random noise with zero mean and unit
variance. We find several values with SNR $>4$, and 1 comb spacing
reaching a SNR of 5.2 (for $\Delta \nu = 61.66$ MHz). However their number
is roughly consistent from what one expects to find by chance when
sampling a gaussian distribution such a large number of times (several times $10^5$
uncorrelated samples). Moreover, we find negative SNR values
approximately as often as positive ones (in particular, we also find a
large negative SNR = -5.7 for $\Delta \nu = 224.22$). As a null test,
we ran the same analysis on the spectrum measured in the off position,
and found a comparable number of comb spacings with SNR $>$ 4, and two
values with SNR $>$ 5 in absolute value.

We therefore conclude that the outcome of our comb search is
consistent with pure noise, and that no PAH combs were detected in our
data within our sensitivity. 

We also searched for the three
corannulene lines within our bandwidth and found an overal SNR consistent
with noise.

A potential caveat of our analysis is that we assumed PAH
lines to be narrower than our resolution. This need not be the case in principle: if the asymmetry parameter and inertial defects are close
to or larger than their critical values, or if broadening by turbulent
motions is significant (with turbulent velocities larger than 5 km/s
corresponding to 0.4 MHz broadening), the lines could be a few
channels wide. However, since, we do not know the exact shape of
these lines, and moreover smoothing in Fourier space reduces the
amplitude of broad features, we have not attempted to search for combs
with broader ``teeth''. 

\subsection{Implications for PAH abundances} \label{sec:bounds}

\subsubsection{Modelling the expected emission}

In order to predict the expected spectrum in the observed bandwidth, we also require a prescription
for the probability distribution $\mathcal{P}(J, K)$ for the
rotational state of the considered PAH. The theoretical framework for
computing this distribution, extending the work of \cite{DL98_long},
is now well developed \citep{spdust1, Hoang_2010, Ysard_2010,
 Hoang_2011, silsbee, spdust_review}. We could therefore in
principle, given the physical conditions in the observed region, and
given the electric dipole moment and asymmetry parameter of any
specific PAH, predict its rotational spectrum. This
prediction could then be used to set upper bounds on the column
density of the considered PAH. Given the large number of
parameters involved and our limited knowledge of their values, however, we have chosen not to take this route in this work.

Instead, and following AH14, we shall rely on the
observed AME spectrum in the Perseus molecular
cloud, and assume that this emission is entirely due to
spinning dust radiation. We shall further assume that any specific PAH
whose abundance we seek to constrain has a rotation state and
dipole moment similar to those of the ``typical'' PAHs responsible for the AME
continuum. This is clearly not strictly correct, as the dipole
moments of different nitrogen-substituted
circumcoronene, for example, vary by a factor as large as 7
\citep{Hudgins_2005}. 
Moreover, the magnitude of the dipole moment strongly affects the power radiated as well
as the peak of the emission (see for example Section 10 of
\citealt{spdust1}). While imperfect, this assumption has however the merit of simplicity.

Assuming the rotational emission of a particular PAH is
an ideal comb when observed with our finite resolution (i.e. assuming
$\delta$ and $\epsilon$ are small enough), we have:
\beq
S_{\nu}^{\rm PAH} \approx \frac{N_{\rm PAH}}{N_{\rm PAH, tot}}
\frac{\Delta \nu_{\rm comb}}{\Delta \nu_{\rm res}}
S_{\nu}^{\rm AME}. \label{eq:estimate}
\eeq
In the above equation, $S_{\nu}^{\rm AME}$ is the continuum AME flux,
which we shall estimate below from existing observations,
$S_{\nu}^{\rm PAH}$ is the flux from a specific PAH in each comb
``tooth'', $N_{\rm PAH}$ is the column density of this PAH, and
$N_{\rm PAH, tot}$ is the total column density of all ``typical'' PAHs
causing the AME. 

The $32$ Jy AME flux measured by the Planck
Collaboration with a resolution of 1.12 degree corresponds (assuming a
uniform flux accross their beam) to a flux $S_{\nu}^{\rm AME} \approx 1.8$ mJy
per 0.5' GBT beam. We shall use this estimate as our fiducial value in
Eq.~(\ref{eq:estimate}). With $\Delta \nu_{\rm res} = 0.3907$ MHz, we
obtain
\beq
S_{\nu}^{\rm PAH} \approx 0.45 ~ \textrm{mJy} ~ \frac{N_{\rm PAH}}{0.001 ~ N_{\rm
    PAH, tot}} \frac{\Delta \nu_{\rm comb}}{100 ~ \rm MHz}. \label{eq:Snu-PAH}
\eeq



\subsubsection{Upper bounds on abundances of specific PAHs}

Given our reached sensitivity $\sigma = 0.4$ mJy, we can set upper bounds on the
abundance of specific PAHs towards Perseus. We do not reach a SNR larger than 5 in our comb search; this implies
that 
\beq
\frac{S_{\nu}^{\rm PAH}}{0.4 ~\rm{mJy}} \sqrt{\frac{\Delta \nu_{\rm
      tot}}{\Delta \nu_{\rm comb}}} \leq 5.
\eeq
Using Eq.~(\ref{eq:Snu-PAH}) to relate the flux in a particular comb
to the fractional abundance of the underlying PAH, decreased by 20\%
to account for filtering in Fourier space, we derive the approximate limit
\beq
\frac{N_{\rm PAH}}{N_{\rm PAH, tot}} \lesssim 0.001 
\sqrt{\frac{100 ~ \rm MHz}{\Delta \nu_{\rm comb}}}.
\eeq
The numerical value of the right-hand side only varies by a factor of
a few between coronene-sized grains ($\Delta \nu_{\rm comb} \approx
340$ MHz) and circumcoronene-sized grains ($\Delta \nu_{\rm comb}
\approx 70$ MHz).

We emphasise that here by specific PAH we mean a molecule uniquely
determined by its matrix of inertia. For example, singly ionized nitrogen-substituted coronene, with
nitrogen in the innermost benzene ring, only $^{12}$C atoms, and no
deuterium substitution, qualifies as a specific PAH. Two forms of nitrogen-substituted coronene,
with the same chemical formula but substitution sites that give rise
to different moments of inertia, would on the other hand be considered as two different specific PAHs. If one wishes to set an upper bound on the
total abundance of, for
example, all derivatives of coronene, we need a prescription for the
fraction of coronene in any particular specific variant. AH14 estimated such
fractions to be $\sim 6$ percent and $\sim 1.6$ percent for coronene
and circumcoronene, respectively, assuming two charge states,
a nitrogen substitution rate (i.e. the ratio of N to C atoms in PAHs)
of 3 per cent, and a $^{12}$C/$^{13}$C ratio of 70. Besides, AH14 assumed negligible rates
of super- or dehydrogenation, neglected possible methyl or ethyl side groups that may be attached
to PAHs \citep{Li_2012}, and assumed a negligible rate of deuterium
substitution. The latter assumption could be significantly inaccurate: \cite{Draine_2006} and \cite{Linsky_2006} argued that PAHs could be heavily
deuterated, with D/H fractions in PAHs possibly reaching several
percent. The quoted fractions of 6 and 1.6 percent should therefore rather
be considered as upper bounds, keeping in mind that the variety within each PAH ``family'' could be significantly larger. 

Using these values, we find an upper
bound on all forms of coronene of approximately 
\beq
N_{\rm coronene, tot} \lesssim 0.01 ~ N_{\rm PAH, tot},
\eeq
and a bound approximately $10$ times weaker for circumcoronene: 
\beq
N_{\rm circumcoronene, tot} \lesssim 0.1 ~ N_{\rm PAH, tot}.
\eeq 

We conclude this section by recalling that many assumptions have been used to derive these bounds, some
rather optimistic:\\
$(i)$ We have assumed that the spectrum of
the searched PAHs peaks around $\sim 20-30$ GHz, like the AME
spectrum; in reality, it could be that quasi-symmetric PAHs have a
significantly larger or smaller dipole moment than the fiducial
population, and that their emission peaks at significantly lower or
higher frequencies. The gain in power due to a larger dipole moment
may not be enough to compensate for the exponential fall of the
distribution function of angular momenta beyond its peak.
\\
$(ii)$ We do not have a good handle of the inertial defect of large
planar PAHs. The comb-like aspect of the rotational emission would be
destroyed if the inertial defect was significantly larger than a few
parts in $10^5$.\\
$(iii)$ We have assumed that the amount of impurities on PAHs is
rather low, and neglected deuterium substitutions and the attachment
of side groups. This biases the conversion from the abundance of a
specific PAH (determined uniquely by its inertia tensor), which is
what is directly constrained here, to the overall abundance of the
``mother PAH'', including all of its derivatives. \\
$(iv)$ We have assumed that turbulent broadening is negligible,
i.e. that turbulent velocities along the line of sight are less than
$\sim 5$ km/s, beyond which lines would spread across several channels.

\section{Conclusion} \label{sec:conclusion}

We have searched for comb-like rotational emission from
quasi-symmetric ``grand PAHs'' in the Perseus molecular cloud, a
region known to harbour AME, most probably caused by spinning PAHs. 
We conducted this search by match-filtering spectroscopic data acquired with the Green
Bank Telescope. Our data reduction method relies on smoothing in
Fourier space and frequency switching in order to eliminate gain fluctuations. We have
reached a noise level of 0.4 mJy per 0.4 MHz channel, over a total
bandwidth of 3 GHz. 

This noise level was not sufficient to make a detection, but allowed
us to set informative upper bounds on the abundance
of specific, quasi-symmetric PAHs, provided the
AME observed in Perseus is primarily due
to small spinning dust grains. In addition, our upper bounds generically apply to a large range of PAHs, with $\sim 15-100$ carbon atoms.

Our upper bounds imply that any specific charge state and substitution
state of a quasi-symmetric PAH cannot make more than $\sim 0.1\%$ of
the the total PAH population. This in turn implies that the total
amount of any quasi-symmetric PAH (i.e.~including all charge and
substitution states) is at most a few percent of the total PAH
population. These bounds should be used with caution, however, as they
rely on assuming that the constrained specific PAHs have similar
properties as the fiducial PAHs producing the spinning dust emission. 

Our non-detection of specific PAHs could very well result from a non-optimal choice of
target or frequency domain, and more searches are warranted, exploring
other regions of the sky with different instruments and at different
wavelengths. Identifying specific PAHs would be an important milestone in our
understanding of the ISM. In this work we have used a new and promising
technique, and hope to have paved the way for more searches to come.

\section*{Acknowledgements}

We are grateful to the GBT staff
for their availability during our visit and to Dan Perera for help
with preliminary data reduction. We also thank Bruce Draine for
detailed comments on this manuscript and Clive Dickinson and Christopher Tibbs
for useful discussions.

Y.A.-H.~was supported by the Frank and Peggy Taplin fellowship while at the
Institute for Advanced study, and is supported by the John
Templeton Foundation award 43770 at the Johns Hopkins University. The National Radio Astronomy Observatory
is a facility of the National Science Foundation operated under
cooperative agreement by Associated Universities, Inc. 

\bibliography{PAHs_GBT_paper}

\end{document}